\documentclass{LMCS}

\usepackage{amsmath,amssymb,graphicx,enumerate,hyperref}

\theoremstyle{remark}\newtheorem{notation}[thm]{Notation}
\theoremstyle{remark}\newtheorem{problem}[thm]{Problem}
\theoremstyle{remark}\newtheorem{remark}[thm]{Remark}
\theoremstyle{plain}\newtheorem{lemma}[thm]{Lemma}
\theoremstyle{plain}\newtheorem{proposition}[thm]{Proposition}

\renewcommand{\emptyset}{\varnothing}

\newcommand{\N}{\mathbb{N}}
\newcommand{\Z}{\mathbb{Z}}

\newcommand{\logic}{\mbox{\rm PTCTL}}
\newcommand{\CTL}{\mbox{\rm CTL}}

\newcommand{\Prog}{\mbox{\rm Run}}
\newcommand{\llambda}{\mbox{\rm Duration}}

\newcommand{\EX}{\exists\!\bigcirc}

\newcommand{\EU}{\exists\mbox{U}}
\newcommand{\AU}{\forall\mbox{U}}
\newcommand{\EG}{\exists\Box}
\newcommand{\AG}{\forall\Box}
\newcommand{\EF}{\exists\Diamond}
\newcommand{\AF}{\forall\Diamond}
\newcommand{\cleq}{\equiv_{a,\leq}}
\newcommand{\coneleq}{\equiv_{1,\leq}}
\newcommand{\cgeq}{\equiv_{a,\geq}}
\newcommand{\ctwogeq}{\equiv_{2,\geq}}
\newcommand{\conegeq}{\equiv_{1,\geq}}
\newcommand{\xconj}{$x$-conjunction}
\newcommand{\tconj}{$\theta$-conjunction}
\newcommand{\true}{\top}
\newcommand{\false}{\bot}
\newcommand{\Alpha}{A}
\newcommand{\pAth}{\leadsto}
\newcommand{\Time}{\mbox{D}}

\newcommand{\duration}{duration}

\newcommand{\cc}{\varsigma}

\def\doi{3 (1:7) 2007}
\lmcsheading%
{\doi}
{1--30}
{}
{}
{Jul.~19, 2005}
{Feb.~27, 2007}
{}

\begin{document}

\title[Parameters Everywhere]{Real-Time Model-Checking:
       Parameters Everywhere\rsuper *}

     \author[V.~Bruy\`ere]{V\'eronique Bruy\`ere\rsuper a} 
     \address{{\lsuper a}Institut d'Informatique, Universit\'e de Mons-Hainaut, 
              Le Pentagone, Avenue du Champ de Mars 6, B-7000 Mons, Belgium.}
       \email{Veronique.Bruyere@umh.ac.be}
     \author[J.-F.~Raskin]{Jean-Fran\c cois Raskin\rsuper b}
     \address{{\lsuper b}D\'epartement d'Informatique, Universit\'e Libre de Bruxelles,
              Boulevard du Triomphe CP 212, B-1050-Bruxelles, Belgium.}
       \email{Jean-Francois.Raskin@ulb.ac.be}

       \thanks{{\lsuper b}This research was supported by the Belgian FNRS grant
         2.4530.02 of the FRFC project ``Centre F\'ed\'er\'e en
         V\'erification."}
       
       \titlecomment{{\lsuper *}A preliminary version of this paper appeared in
         the Proceedings of the 23rd Conference on Foundations of
         Software Technology and Theoretical Computer Science,
         FSTTCS'03, {\it Lecture Notes in Computer Science} 2914,
         Springer, 2003, pp. 100-111 (see \cite{prelim}).}

\begin{abstract}
  In this paper, we study the model-checking and parameter synthesis problems
  of the logic TCTL over discrete-timed automata where parameters are allowed
  both in the model (timed automaton) and in the property (temporal formula).
  Our results are as follows. On the negative side, we show that the
  model-checking problem of TCTL extended with parameters is undecidable over
  discrete-timed automata with only one parametric clock. The undecidability
  result needs equality in the logic. On the positive side, we show that the
  model-checking and the parameter synthesis problems become decidable for a
  fragment of the logic where equality is not allowed. Our method is based on
  automata theoretic principles and an extension of our method to express
  durations of runs in timed automata using Presburger arithmetic.
\end{abstract}

\maketitle

\section{Introduction}

In this paper, we further investigate the model-checking problem of real-time
formalisms with parameters. In recent works, parametric real-time
model-checking problems have been studied by several authors. 

Alur et al study in~\cite{AHV93} the analysis of discrete- and dense-timed
automata where clocks are compared to parameters. For this class of parametric
timed automata, they focus on the emptiness problem: are there concrete values
for the parameters so that the automaton has an accepting run? They show that
when only one clock is compared to parameters, the emptiness problem is
decidable. But this problem becomes undecidable when three clocks are compared
to parameters.\footnote{The authors mention the case of two clocks as an open
  problem.} Hune et al study in \cite{HRSV} a subclass of parametric
dense-timed automata (L/U automata) such that each parameter occurs either as
a lower bound or as an upper bound.

Wang in~\cite{wang95,wang97parametric}, Emerson et al
in~\cite{emerson99parametric}, Alur et al in~\cite{alur99parametric} and the
authors of this paper in~\cite{VJFstacs03} study the introduction of
parameters in temporal logics. The model-checking problem for TCTL extended
with parameters over discrete- and dense-timed automata (without parameters)
is decidable. On the other hand, only a fragment of LTL extended with
parameters is decidable.

Unfortunately, in all those previous works, the parameters are {\em only} in
the model (expressed as a timed automaton) or {\em only} in the property
(expressed as a temporal logic formula). Nevertheless, when expressing a
temporal property of a parametric system, it is {\em natural} to refer in the
temporal formula to the parameters used in the system.

In this paper, we study the model-checking problem of the logic TCTL {\em
  extended with parameters} over the runs of a {\em discrete}-timed automaton
with {\em one parametric clock}. To the best of our knowledge, this is the
first work that studies the model-checking and parameter synthesis problems
with parameters both in the model and in the property. We restrict to one
parametric clock since the emptiness problem for discrete-time automata with
three parametric clocks is already undecidable (see above, \cite{AHV93}). The
case of dense-timed automata with one parametric clock is not investigated in
this paper.

\begin{figure}
        \begin{center}
                \includegraphics[scale=.50]{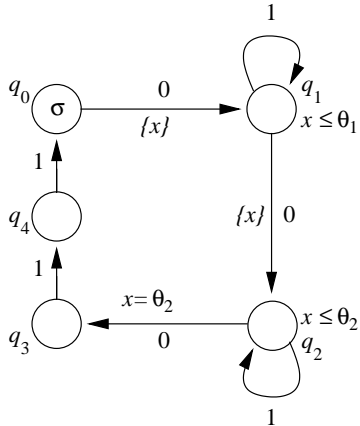}
        \end{center}
        \caption{A parametric timed automaton}
        \label{fIg:intro}
\end{figure}

Let us illustrate the kind of properties that we can express with a parametric
temporal logic over a parametric timed automaton. The automaton ${\sf A}$ of
Figure~\ref{fIg:intro}
is a discrete-timed automaton with one clock $x$ and two parameters $\theta_1$
and $\theta_2$. Here we explicitly model the elapse of time by transitions
labeled by $0$ or $1$. State $q_0$ is labeled with atomic proposition $\sigma$
and in all other states this proposition is false. The possible runs of this
automaton starting at $q_0$ are as follows. The control instantaneously leaves
$q_0$ and goes through $q_1,q_2,q_3$ to come back in $q_0$, the time spent in
this cycle is constrained by the parameters $\theta_1$ and $\theta_2$. In
fact, the control has to leave $q_1$ at most $\theta_1$ time units after
entering it and the control has to stay exactly $\theta_2$ time units in state
$q_2$. To express properties of those behaviors, we use TCTL logic augmented
with parameters. Let us consider the next three formulae for configuration
$(q_0,0)$, i.e. the control is in state $q_0$ and clock $x$ has value $0$:

\begin{enumerate}[(i)]
  \item $\AG (\sigma \rightarrow \AF_{\leq \theta_3} \sigma)$
  \item $\forall \theta_1 \forall \theta_2 \cdot (\theta_2 \leq
    \theta_1 \rightarrow \AG (\sigma \rightarrow \AF_{\leq 2\theta_1 + 2}
    \sigma))$
\item $\forall \theta_1 \cdot (\theta_1 \geq 5 \rightarrow \AG
  (\sigma \rightarrow \AF_{< 2\theta_1 + 2} \sigma))$
\end{enumerate}

\noindent 
The {\em parameter synthesis problem} associated to formula $(i)$, asks for
which values of $\theta_1,\theta_2$ and $\theta_3$, the formula is {\sc true}
at configuration $(q_0,0)$. By observing the model and the formula, we can
deduce the following constraint on the parameters: $\theta_3 \geq \theta_1 +
\theta_2 + 2$. This means that any cycle through the four states has duration
bounded by $\theta_1 + \theta_2 + 2$. Formula $(ii)$ formalizes the next
question ``In all the cases where the value assigned to parameter $\theta_1$
is greater than the value assigned to parameter $\theta_2$, is it true that
any cycle has a duration bounded by $2 \theta_1 + 2$''. As there is no free
parameter in the question, the question has a {\sc yes-no} answer. This is a
{\em model-checking problem}. For formula $(ii)$, the answer is {\sc yes} in
configuration $(q_0,0)$. Finally, formula $(iii)$ lets parameter $\theta_2$
free and formalizes the question ``What are the possible values that can be
given to $\theta_2$ such that for any value of $\theta_1 \geq 5$, a cycle
through the four states lasts at most $2\theta_1+1$ time units''. This is
again a parameter synthesis problem and the answer is $\theta_2 \leq 4$.

In this paper, we study the algorithmic treatment of such problems. Our
results are as follows. On the negative side, we show that the model-checking
problem of TCTL extended with parameters is {\em undecidable} over timed
automata with {\em only one} parametric clock. The undecidability result needs
{\em equality in the logic}. On the positive side, we show that the
model-checking problem becomes {\em decidable} and the parameter synthesis
problem is solvable for a fragment of the logic where the equality is not
allowed. Our algorithm is based on automata theoretic principles and an
extension of our method (see \cite{VJFstacs03}) to express durations of runs
in a timed automaton using Presburger arithmetic. As a corollary, we obtain
the decidability of the emptiness problem for discrete-timed automata with one
parametric clock proved by Alur et al in~\cite{AHV93}. All the formulae given
in the example above are in the decidable fragment.

The paper is organized as follows. In Section 2, we introduce the model of one
parametric clock discrete-timed automaton and the parametric extension of TCTL
that we consider. In Section 3, we establish the undecidability of the
model-checking problem if equality can be used in the logic and we show how to
solve the problem algorithmically for a fragment of the logic where equality
is not allowed.  Proofs of two important propositions introduced in Section 3
are postponed to Section 4. We finish the paper in Section 5 by drawing some
conclusions.

\section{Parameters Everywhere}\label{sec}

In this section, we introduce \emph{parameters} in the automaton used to model
the system \emph{as well as} in the logic used to specify properties of the
system. The automata are parametric timed automata as defined in \cite{AHV93}
with a \emph{discrete} time domain and \emph{one} parametric clock. The logic
is Parametric Timed CTL Logic as defined in \cite{VJFstacs03}. We introduce
the problems that we want to solve and we conclude the section with an
example.

\begin{notation}        \label{not:guard}
  Let $\Theta$ be a fixed finite set of \emph{parameters} $\theta$ that are
  \emph{shared} by the automaton and the logical formulae. A \emph{parameter
    valuation} for $\Theta$ is a function $v : \Theta \rightarrow \N$ which
  assigns a natural number to each parameter $\theta \in \Theta$. In the
  sequel, $\alpha, \beta, \ldots$ mean any linear term $\Sigma_{i \in I} c_i
  \theta_i + c$, with $c_i, c \in \N$ and $\{\theta_i | i \in I \} \subseteq
  \Theta$. A parameter valuation $v$ is naturally extended to linear terms by
  defining $v(c) = c$ for any $c \in \N$.

\noindent
We denote by $x$ the \emph{unique} parametric clock. The same notation $x$ is
used for both the clock and a value of the clock. A \emph{guard} $g$ is any
conjunction of $x \sim \alpha$ with $\sim \; \in \{ = , < , \leq , > , \geq
\}$. We denote by ${\sf G}$ the set of guards. Notation $x \models_v g$ means
that $x$ satisfies $g$ under valuation $v$. We use notation $\Sigma$ for the
set of \emph{atomic propositions}.
\end{notation}

\subsection{Parametric Timed Automata}

We recall the definition of one parametric clock discrete-timed automata as
introduced in \cite{AHV93}.

We make the hypothesis that non-parametric clocks have all been suppressed by
a technique related to the region construction, see~\cite{AHV93} for details.

\begin{defi} \label{def:timed_aut}
  A \emph{parametric timed automaton} ${\sf A}$ is a tuple $(Q,E,{\sf
    L},{\sf I})$, where $Q$ is a finite set of \emph{states}, $E \subseteq Q
  \times \{0,1\} \times {\sf G} \times 2^{\{x\}} \times Q$ is a finite set of
  \emph{edges}, ${\sf L} : Q \rightarrow 2^{\Sigma}$ is a \emph{labeling}
  function and ${\sf I} : Q \rightarrow {\sf G}$ assigns an \emph{invariant}
  ${\sf I}(q) \in {\sf G}$ to each state $q$.

\noindent
A \emph{configuration} of ${\sf A}$ is a pair $(q,x)$, where $q$ is a state
and $x$ is a clock value.
\end{defi}

Whenever a parameter valuation $v$ is given, ${\sf A}$ becomes a usual
one-clock timed automaton denoted by ${\sf A}^v$. We recall the next
definitions of transition and run in ${\sf A}^v$.

\begin{defi} \label{def:config}
  Let $v$ be a parameter valuation. A \emph{transition} $(q,x)
  \stackrel{\tau}\rightarrow (q',x')$ between two configurations $(q,x)$ and
  $(q',x')$, with time increment $\tau \in \{0,1\}$, is allowed in ${\sf
    A}^v$ if (1) $x \models_v {\sf I}(q)$ and $x' \models_v {\sf I}(q')$,
  (2) there exists an edge $(q,\tau,g,r,q') \in E$ such that $x + \tau
  \models_v g$ and $x' = 0$ if $r = \{x\}$, $x' = x + \tau$ if $r =
  \emptyset$.\footnote{Note that time increment $\tau$ is first added to $x$,
    guard $g$ is then tested, and finally $x$ is reset according to~$r$.}

A \emph{run} $\rho = (q_i,x_i)_{i \geq 0}$ of ${\sf A}^v$ is an infinite
sequence of transitions $(q_i,x_i) \stackrel{\tau_i}\rightarrow
(q_{i+1},x_{i+1})$ such that $\Sigma_{i \geq 0} \tau_i = \infty$.\footnote{Non
  Zenoness property.} The \emph{\duration} $t = \Time_{\rho}(q_i,x_i)$ at
configuration $(q_i,x_i)$ of $\rho$ is equal to $t = \Sigma_{0 \leq j < i}
\tau_{j}$. A \emph{finite run} $\rho$ is a finite sequence of transitions. It
is shortly denoted by $(q,x) \pAth (q',x')$ such that $(q,x)$ (resp.
$(q',x')$) is its first (resp. last) configuration. Its \emph{duration}
$\Time_{\rho}$ is equal to $\Time_{\rho}(q',x')$.
\end{defi}

\subsection{Parametric Timed CTL Logic} \label{sec:PTCTL}

Formulae of \emph{Parametric Timed CTL logic}, \logic\ for short, are formed
by a block of quantifiers over some parameters followed by a quantifier-free
temporal formula. They are defined as follows. Notation $\sigma$ means any
atomic proposition $\sigma \in \Sigma$ and $\alpha, \beta$ are linear terms as
before.

\begin{defi}      \label{def:syntax}
A \logic\ formula $f$ is of the form
$$f = Q_1\theta_1 \; \cdots \; Q_k \theta_k \; \varphi$$
such that $k \geq 0$,
$\{\theta_1, \ldots, \theta_k \} \subseteq \Theta$, $Q_j \in \{\exists,
\forall\}$ for each $j$, $1 \leq j \leq k$, and $\varphi$ is given by the
following grammar
$$\varphi ::= \sigma ~|~ \alpha \sim \beta ~|~ \neg \varphi ~|~ \varphi \vee
\varphi ~|~ \EX \varphi ~|~ \varphi \EU_{\sim\alpha} \varphi ~|~ \varphi
\AU_{\sim\alpha} \varphi$$
\end{defi}

Note that usual operators $\EU$ and $\AU$ are obtained as $\EU_{\geq
0}$ and $\AU_{\geq 0}$.  We also use the following abbreviations:
$\EF_{\sim \alpha} \varphi$ for $\top \EU_{\sim \alpha} \varphi$,
$\AF_{\sim \alpha} \varphi$ for $\top \AU_{\sim \alpha} \varphi$,
$\EG_{\sim \alpha} \varphi$ for $\neg \AF_{\sim \alpha} \neg \varphi$,
and $\AG_{\sim \alpha} \varphi$ for $\neg \EF_{\sim \alpha} \neg
\varphi$.

We use notation QF-\logic\ for the set of \emph{quantifier-free} formulae
$\varphi$ of \logic. The set of parameters of $\Theta$ that are \emph{free} in
$f$, that is, not under the scope of a quantifier, is denoted by $\Theta_f$.
Thus, for a QF-\logic\ formula $\varphi$, we have $\Theta_{\varphi} = \Theta$
(recall that $\Theta$ is the set of parameters that appear in the formula
\emph{and} in the automaton).

We now give the \emph{semantics} of \logic.

\begin{defi}      \label{def:semantics}
  Let $\sf A$ be a parametric timed automaton and $(q,x)$ be a configuration
  of $\sf A$. Let $f = Q_1\theta_1 \; \cdots \; Q_k \theta_k \; \varphi$ be a
  \logic\ formula. Given a parameter valuation $v$ on $\Theta_f$, the
  \emph{satisfaction} relation $(q,x) \models_v f$ is defined inductively as
  follows. If $f = \varphi$, then $(q,x) \models_v \varphi$ according to the
  following rules:
  \begin{enumerate}[$\bullet$]
  \item $(q,x)\models_v \sigma$ iff there exists\footnote{We verify the
      existence of a run starting in $(q,x)$ to ensure that time can progress
      in ${\sf A}^v$ from that configuration.} a run $\rho = (q_i,x_i)_{i
      \geq 0}$ in ${\sf A}^v$ with $(q,x) = (q_0,x_0)$ and $\sigma \in {\sf
      L}(q)$
        \item $(q,x)\models_v \alpha \sim \beta$ iff there exists a run $\rho
          = (q_i,x_i)_{i \geq 0}$ in ${\sf A}^v$ with $(q,x) = (q_0,x_0)$ and
          $v(\alpha) \sim v(\beta)$
        \item $(q,x)\models_v \neg \varphi$  iff $(q,x) \not\models_v \varphi$
        \item $(q,x)\models_v \varphi \vee \psi$ iff $(q,x) \models_v \varphi$
          or $(q,x) \models_v \psi$
        \item $(q,x)\models_v \EX \varphi$ iff there exists a run $\rho =
          (q_i,x_i)_{i \geq 0}$ in ${\sf A}^v$ with $(q,x) = (q_0,x_0)$ and
          $(q_1,x_1) \models_v \varphi$
        \item $(q,x)\models_v \varphi \EU_{\sim \alpha} \psi$ iff there exists
          a run $\rho = (q_i,x_i)_{i \geq 0}$ in ${\sf A}^v$ with $(q,x) =
          (q_0,x_0)$, there exists $i \geq 0$ such that $\Time_{\rho}(q_i,x_i)
          \sim v(\alpha)$, $(q_i,x_i) \models_v \psi$ and $(q_j,x_j) \models_v
          \varphi$ for all $j < i$
        \item $(q,x)\models_v \varphi \AU_{\sim \alpha} \psi$ iff for any run
          $\rho = (q_i,x_i)_{i \geq 0}$ in ${\sf A}^v$ with $(q,x) =
          (q_0,x_0)$, there exists $i \geq 0$ such that $\Time_{\rho}(q_i,x_i)
          \sim v(\alpha)$, $(q_i,x_i) \models_v \psi$ and $(q_j,x_j) \models_v
          \varphi$ for all $j < i$
  \end{enumerate}
  If $f = \exists \theta f'$, then $(q,x) \models_v f$ iff there exists $c \in
  \N$ such that $(q,x) \models_{v'} f'$ where $v'$ is defined on $\Theta_{f'}$
  by $v' = v$ on $\Theta_f$ and $v'(\theta)= c$. If $f = \forall \theta f'$,
  then $(q,x) \models_v f$ iff for all $c \in \N$, $(q,x) \models_{v'} f'$
  where $v'$ is defined on $\Theta_{f'}$ by $v' = v$ on $\Theta_f$ and
  $v'(\theta)= c$.
\end{defi}

\subsection{Problems}

The problems that we want to solve in this paper are the following ones. The
first problem is the model-checking problem for \logic\ formulae $f$ with
\emph{no} free parameters. In this case, we omit the index by $v$ in the
satisfaction relation $(q,x) \models f$ since no parameter (neither in the
automaton nor in the formula) has to receive a valuation.

\begin{problem}         \label{prob:model-checking}
  The \emph{model-checking} problem is the following. Given a parametric timed
  automaton $\sf A$ and a \logic\ formula $f$ such that $\Theta_f =
  \emptyset$, given a configuration $(q,x)$ of $\sf A$, does $(q,x) \models
  f$ hold ?
\end{problem}

The second problem is the more general problem of parameter synthesis for
\logic\ formulae $f$ such that $\Theta_f$ is \emph{any} subset of $\Theta$.

\begin{problem}         \label{prob:parameter-synthesis}
  The \emph{parameter synthesis} problem is the following. Given a parametric
  timed automaton $\sf A$ and a configuration $(q,x)$ of $\sf A$, given a
  \logic\ formula $f$, compute a symbolic representation\footnote{For instance
    this representation could be given in a decidable logical formalism.} of
  the set of parameter valuations $v$ on $\Theta_f$ such that $(q,x) \models_v
  f$.
\end{problem}

\noindent\medskip
{\bf Example} We consider the example given in the introduction with the
parametric timed automaton $\sf A$ of Figure \ref{fIg:intro} and the two
\logic\ formulae respectively equal to
$$f : \forall \theta_1 \forall \theta_2 \cdot (\theta_2 \leq \theta_1
\rightarrow \AG (\sigma \rightarrow \AF_{\leq 2\theta_1 + 2} \,\sigma))$$
and
$$g : \forall \theta_1 \cdot (\theta_1 \geq 5 \rightarrow \AG (\sigma
\rightarrow \AF_{< 2\theta_1 + 2} \,\sigma)).$$
Then $\Theta = \{ \theta_1,
\theta_2 \}$, $\Theta_f = \emptyset$ and $\Theta_g = \{\theta_2\}$. The
model-checking problem ``does $(q_0,0) \models f$ hold" has a {\sc yes}
answer. The parameter synthesis problem ``for which parameter valuations $v$
on $\Theta_g$ does $(q_0,0) \models_v g$ hold" receives the answer $\theta_2
\leq 4$.

\subsection{Comments} \label{com}

We end Section~\ref{sec} by some comments on the definitions and the problems
presented above.

\begin{enumerate}[(1)]
\item We consider timed automata with only one parametric clock for the
  following reason. In~\cite{AHV93}, the authors investigate the following
  emptiness problem, which is a particular case of
  Problem~\ref{prob:model-checking} : are there concrete values for the
  parameters so that a parametric timed automaton has an accepting run? They
  show that the emptiness problem is decidable when there is one parametric
  clock, that this problem is open for two parametric clocks, and that it
  becomes undecidable for three parametric clocks. They illustrate the
  hardness of the two-clock emptiness problem by presenting connections with
  difficult open problems in logic and automata theory.
  
  Both discrete time and dense time are considered in~\cite{AHV93} (see
  \cite{miller} for further results), whereas we only deal with discrete time
  in this paper.

\item To solve Problem~\ref{prob:model-checking}, we use the same approach as
  in our paper \cite{VJFstacs03} where we propose a simple proof of the
  model-checking problem for \logic\ over timed-automata without parameters.
  We prove in \cite{VJFstacs03} that the durations of runs starting from a
  region and ending in another region can be defined by a formula of
  Presburger arithmetic. It follows that the model-checking problem can be
  reduced to checking whether some sentence of Presburger arithmetic is true or
  false.
  
  This approach is different from the one used in \cite{ACD90} when there is
  no parameter at all. We recall that in \cite{ACD90}, an extra clock is added
  to the timed automaton and the model-checking is solved thanks to a labeling
  (like for $\CTL$) of the region graph of the augmented automaton. We have
  not investigated this kind of approach here, because the additional clock
  would be parametric, leading to two parametric clocks inside the automaton.
  
\item Linear terms $\alpha$ are present in the definition of parametric timed
  automata (inside the guards and the invariants) as well as in the definition
  given for $\logic$. More generally full Presburger arithmetic is present in
  $\logic$. Alternative restricted definitions could be
\begin{enumerate}[$\bullet$]
\item for parametric timed automata : guards and invariants are restricted to
  conjunctions of $x \sim \theta$, $x \sim c$ (instead of any conjunction of
  $x \sim \alpha$);
\item for $\logic$ : the restricted grammar
$$\varphi ::= \sigma ~|~ \neg \varphi ~|~ \varphi \vee \varphi ~|~ \EX
  \varphi ~|~ \varphi \EU_{\sim\theta} \varphi ~|~ \varphi \EU_{\sim c}
  \varphi ~|~ \varphi \AU_{\sim\theta} \varphi ~|~ \varphi \AU_{\sim c} \varphi
$$
is used instead of the grammar proposed in Definition~\ref{def:syntax}. 
\end{enumerate}
In this way, the constraints over the parameters are restricted to comparisons
with a parameter or with a constant, instead of comparisons with a linear term
over parameters.
  
However we observe in Remark~\ref{rem} below that the undecidability result
about the model-checking problem is the same when using
Definitions~\ref{def:timed_aut} and~\ref{def:syntax}, or with the above
restricted definitions.
\end{enumerate}

\section{Decision Problems}

In this section, we prove that the model-checking problem is undecidable. The
undecidability comes from the use of equality in the operators $\EU_{\sim
  \alpha}$ and $\AU_{\sim \alpha}$. Then for a fragment F-\logic\ of \logic\ 
where equality is forbidden, we prove that the model-checking problem becomes
decidable. In this case, we also positively solve the parameter synthesis
problem. Our proofs use Presburger arithmetic and its extension with integer
divisibility.

Let us introduce the precise definition of the fragment F-\logic.\footnote{In
  the preliminary version \cite{prelim} of this paper, we considered a
  fragment of \logic\ that is larger than F-\logic. The grammar of the
  proposed fragment was equal to the grammar proposed in
  Definition~\ref{def:fragment} extended with $\varphi \AU_{> \alpha} \varphi$
  and $\varphi \AU_{\geq \alpha} \varphi$. 
We have found a mistake in the proof of the decidability of the model-checking
for this fragment.}

\begin{defi} \label{def:fragment}
  Notation F-\logic\ is used to denote the fragment of \logic\ where the
  equality is forbidden in the operators $\EU_{\sim \alpha}$ and $\AU_{\sim
    \alpha}$ and the inequalities $>, \geq$ are forbidden in $\AU_{\sim
    \alpha}$. More precisely, a F-\logic\ formula $f$ is of the form $f =
  Q_1\theta_1 \; \cdots \; Q_k \theta_k \; \varphi$ such that $\varphi$ is
  given by the grammar
  \begin{eqnarray*} 
\varphi &::=& \sigma ~|~ \alpha \sim \beta ~|~ \neg \varphi ~|~ \varphi \vee
  \varphi ~|~ \EX \varphi \\
 & &~|~\varphi \EU_{< \alpha} \varphi ~|~ \varphi
  \EU_{\leq \alpha} \varphi ~|~\varphi \EU_{> \alpha} \varphi ~|~\varphi
  \EU_{\geq \alpha} \varphi \\
 & &~|~\varphi \AU_{< \alpha} \varphi ~|~ \varphi \AU_{\leq \alpha} \varphi
~|~ \varphi \AU \varphi
\end{eqnarray*}
\end{defi}

\subsection{Undecidability for PTCTL}

We prove here that Problem \ref{prob:model-checking} is undecidable for
\logic. The proof relies on the undecidability of Presburger arithmetic with
divisibility.

{\em Presburger arithmetic with divisibility} is an extension of Presburger
arithmetic with integer divisibility relation. The additional divisibility
relation is denoted by $z | z'$ and means ``$z$ divides $z'$''. Every formula
of Presburger arithmetic with divisibility can be put into \emph{normal form}:
\begin{eqnarray} \label{eq}
Q z_1 Q z_2 \dots Q z_n ~ (\neg) \phi_1 \star (\neg) \phi_2 \star \dots
\star (\neg) \phi_m
\end{eqnarray}
where $\star$ belongs to $\{ \lor, \land \}$, $(\neg)$
means that negation is optional and each $\phi_i$ is one of the following
atomic formulae: $(i)$ $z = \alpha$, $(ii)$ $z > \alpha$, $(iii)$ $z | z'$
such that $\alpha$ is a linear term and $z' > 0$. While Presburger arithmetic
has a decidable theory, Presburger arithmetic with divisibility is
undecidable~\cite{bes}.

\begin{thm} \label{thm:undecidable}
  For any sentence $\Phi$ of Presburger arithmetic with divisibility, we can
  construct a parametric timed automaton ${\sf A}$, a configuration $(q,x_0)$
  and a \logic\ formula $f$ such that $\Phi$ is {\sc true} iff the
  answer to the model-checking problem $(q,x_0) \models f$ for $\sf A$ is
  {\sc yes}.
\end{thm}

\proof
  Let us make the assumption that the sentence $\Phi$ is in normal form
  (\ref{eq}). We are going to construct a \logic\ formula $f$ and a
  parametric timed automaton $\sf A$. The set $\Theta$ of parameters is equal
  to the set of all the variables used in $\Phi$.
  
\begin{figure}
        \begin{center}
                \includegraphics[scale=.55]{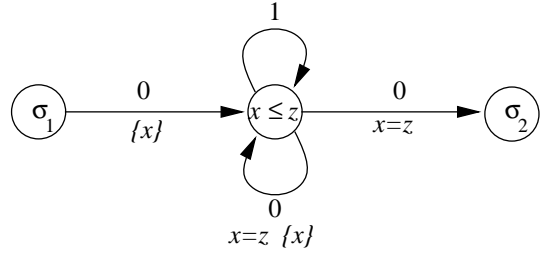}
        \end{center}
        \caption{Automaton for $z | z'$}
        \label{fIg:div}
\end{figure}

  For each subformula $\phi_l$ of the form $z = \alpha$ or $z > \alpha$, we
  define the \logic\ formula $\hat{\phi}_l$ equal to $\phi_l$. For each
  subformula $\phi_l$ of the form $z | z'$, we construct the next parametric
  timed automaton ${\sf A}_{\phi_l}$ and \logic\ formula $\hat{\phi}_l$.
  The automaton ${\sf A}_{\phi_l}$ is given in Figure~\ref{fIg:div}.
We label the unique initial state $i_l$ of this automaton by $\sigma^l_1$ and
the unique final $f_l$ state by $\sigma^l_2$. It is easy to see that there is
a run $\rho$ from the initial configuration $(i_l,0)$ to the final
configuration $(f_l,z)$ with duration $D_{\rho}$ iff $z | D_{\rho}$. For
formula $\hat{\phi}_l$, we take $\sigma^l_1 \land \EF_{=z'} \sigma^l_2$.

Now we construct formula $f$ as follows
$$f ~:~ Q z_1 Q z_2 \cdots Q z_n ~ (\neg) \hat{\phi}_1 \star (\neg)
\hat{\phi}_2 \star \dots \star (\neg) \hat{\phi}_m.$$
We construct the
automaton ${\sf A}$ by first taking the union of all the previous automata
${\sf A}_{\phi_l}$ (introduced for the divisibility subformulae). We then
merge their initial states into a unique state of ${\sf A}$ that we call $q$.
The label ${\sf L}(q)$ of $q$ is the union of the labels $\sigma^l_1$.
Finally, we add a new state $q'$ to ${\sf A}$ and an edge
$(f_l,0,\true,\emptyset,q')$ from any final state $f_l$ of ${\sf A}_{\phi_l}$
to state $q'$ labeled with $\tau = 0$ and without any guard and reset. To
complete the construction, we add a self-loop $(q',1,\true,\emptyset,q')$ on
$q'$ that allows time to progress.

It is easy to see that given ${\sf A}$, we have $(q,0) \models f$ iff $\Phi$
is {\sc true}.
\qed

As a direct consequence of Theorem \ref{thm:undecidable}, we have:

\begin{cor}       \label{cor:undecidable}
The model-checking problem for \logic\ is undecidable.
\end{cor}

\begin{remark}
  In the previous proof, all the proposed \logic\ formulae $\hat{\phi}_l$ only
  use the subscript $=$ in the operators $\EU_{\sim\theta}$ and
  $\AU_{\sim\theta}$. It follows that the model-checking problem is already
  undecidable with the grammar 
$$\varphi ::= \sigma ~|~ \alpha \sim \beta ~|~ \neg \varphi ~|~ \varphi \vee
  \varphi ~|~ \EX \varphi ~|~ \varphi \EU_{=\alpha} \varphi ~|~ \varphi
  \AU_{=\alpha} \varphi$$
instead of the grammar given in Definition~\ref{def:syntax}.
\end{remark}

\begin{remark} \label{rem}
  Given a sentence $\Phi$ of Presburger arithmetic with divisibility, we have
  shown in the proof of Theorem~\ref{thm:undecidable} how to construct a
  parametric timed automaton ${\sf A}$, a configuration $(q,x_0)$ and a
  \logic\ formula $f$ such that $\Phi$ is {\sc true} iff the answer to the
  model-checking problem $(q,x_0) \models f$ for $\sf A$ is {\sc yes}.
  
  As mentioned in Section~\ref{com} (see Comment 3), we could consider
  alternative restricted definitions for parametric timed automata and
  $\logic$. We say that a parametric timed automaton is \emph{restricted} and
  that a formula of $\logic$ is \emph{restricted} if they respect the
  restricted definitions given in Comment 3 of Section~\ref{com}.
  
  Let us show that given a sentence $\Phi$ of Presburger arithmetic with
  divisibility, we can construct a restricted parametric timed automaton ${\sf
    A}$, a configuration $(q,x_0)$ and a restricted formula $f$ of $\logic$
  such that $\Phi$ is {\sc true} iff the answer to the model-checking problem
  $(q,x_0) \models f$ for $\sf A$ is {\sc yes}.  The proof is in the same vein
  as the previous one. The sentence $\Phi$ is supposed to be in normal form
  like in (\ref{eq}) with each subformula $\phi_l$ of the form $z = \alpha$,
  $z > \alpha$, or $z | z'$. We first treat the case $z = \alpha$ (with hints
  on the construction with $\alpha = 2 \theta + 2$). Instead of defining
  $\hat{\phi}_l$ equal to $\phi_l$ as in the previous proof, we consider the
  restricted parametric timed automaton of Figure~\ref{fIg:presburger}, and
  the restricted formula $\hat{\phi}_l$ equal to $\sigma^l_1 \land
  \EF_{=z}\sigma^l_2$.
\begin{figure}
        \begin{center}
                \includegraphics[scale=.55]{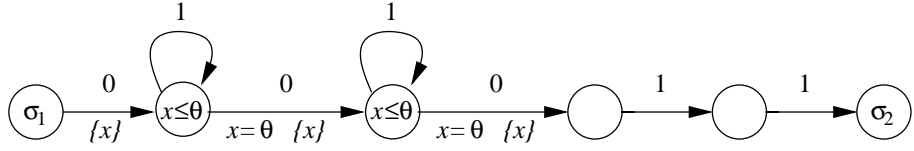}
        \end{center}
        \caption{Automaton for $z = 2 \theta + 2$}
        \label{fIg:presburger}
\end{figure}
The case $z > \alpha$ is treated similarly : for the example of $z > 2\theta +
1$, the automaton is the one of Figure~\ref{fIg:presburger} with an additional
loop with label 1 on the rightmost location, and the formula is again equal
to $\sigma^l_1 \land \EF_{=z}\sigma^l_2$. Finally the case $z | z'$ is treated
as in the previous proof since the automaton and the formula that were
proposed are both restricted.

It follows that the model-checking problem with the restricted definitions of
parametric timed automata and logic \logic\ is still undecidable. Notice that
again all the proposed restricted formulae $\hat{\phi}_l$ only use the
equality in the operators $\EU_{\sim\theta}$ and $\AU_{\sim\theta}$.


\end{remark}

\subsection{Decidability for F-PTCTL} \label{sec:general}

In this section, we provide solutions to the model-checking problem and the
parameter synthesis problem for F-\logic. Our approach is as follows. Given a
state $q$ and a formula $\varphi$ of QF-F-\logic\footnote{Notation QF- has
  been introduced after Definition~\ref{def:syntax} to mention that $\varphi$
  is a quantifier free formula.}, we construct a
Presburger formula $\Delta_{q,\varphi}(x,\Theta)$ with $x$ and all $\theta \in
\Theta$ as free variables such that
$$(q,x_0) \models_v \varphi \quad\mbox{iff}\quad
\Delta_{q,\varphi}(x_0,v(\Theta)) \mbox{ is {\sc true}}$$
for \emph{any}
valuation $v$ on $\Theta$ and \emph{any} value $x_0$ of the clock (see Theorem
\ref{thm:main}). Solutions to Problems \ref{prob:model-checking} and
\ref{prob:parameter-synthesis} will be obtained as a corollary (see
Corollaries \ref{cor:1} and \ref{cor:2}). For instance, the decidability of
the model-checking problem will derive from the decidability of Presburger
arithmetic. Indeed, if we denote by $Q \Theta \; \varphi$ a F-\logic\ formula
$f$ with no free parameters, then to test if $(q,x_0) \models f$ is equivalent
to test if the sentence $Q \Theta \; \Delta_{q,\varphi}(x_0,\Theta)$ is {\sc
  true}.

\medskip\noindent {\bf Example} Consider the parametric timed automaton of
Figure \ref{fIg:intro} and the QF-F-\logic\ formula $\varphi$ equal to $\AG
(\sigma \rightarrow \AF_{\leq \theta_3} \sigma)$. Then $\Theta =
\{\theta_1,\theta_2,\theta_3\}$. Presburger formula
$\Delta_{q_0,\varphi}(x,\Theta)$ is here equal to $\theta_1 + \theta_2 + 2
\leq \theta_3$ with no reference to $x$ since it is reset along the edge from
$q_0$ to $q_1$. Thus $(q,x_0) \models_v \varphi$ for any clock value $x_0$ and
any valuation $v$ such that $v(\theta_1) + v(\theta_2) + 2 \leq v(\theta_3)$.
The model-checking problem $(q,x_0) \models \forall \theta_1 \forall \theta_2
\exists \theta_3 \varphi$ has a {\sc yes} answer for any $x_0$ because the
sentence $\forall \theta_1 \forall \theta_2 \exists \theta_3 ~ (\theta_1 +
\theta_2 + 2 \leq \theta_3)$ is {\sc true} in Presburger arithmetic. If clock
$x$ was not reset along the edge from $q_0$ to $q_1$, then the formula
$\Delta_{q_0,\varphi}(x,\Theta)$ would be equal to $(\theta_1 + \theta_2 + 2
\leq \theta_3) \wedge (x \leq \theta_1)$ and the above model-checking problem
would have a {\sc yes} answer iff $\forall \theta_1 \forall \theta_2 \exists
\theta_3 ~ (\theta_1 + \theta_2 + 2 \leq \theta_3) \wedge (x_0 \leq
\theta_1)$, that is $x_0 = 0$.

\medskip

As indicated by this example, the Presburger formula
$\Delta_{q,\varphi}(x,\Theta)$ constructed from the QF-F-\logic\ formula
$\varphi$ is a boolean combination of terms of the form $\theta \sim \alpha$
or $x \sim \alpha$ where $\theta$ is a parameter, $x$ is the clock and
$\alpha$ is a linear term over parameters. Formula
$\Delta_{q,\varphi}(x,\Theta)$ must be seen as a \emph{syntactic} translation
of formula $\varphi$ into Presburger arithmetic. The question ``does $(q,x_0)
\models f$ hold" with $f = Q \Theta \; \varphi$ is translated into the
question ``is the Presburger sentence $Q \Theta \;
\Delta_{q,\varphi}(x_0,\Theta)$ {\sc true}". At this point only,
\emph{semantic} inconsistencies inside $Q \Theta \;
\Delta_{q,\varphi}(x_0,\Theta)$ are looked for to check if this sentence is
{\sc true} or not.

\medskip

Our proofs require to work with a set $\sf G$ of guards that is more general
than in Notation \ref{not:guard}.

\begin{notation}        \label{not:ext_guard}
  Linear terms $\alpha, \beta, \ldots$ are any $\Sigma_i c_i \theta_i + c$,
  with $c_i, c \in \Z$ (instead of $\N$). Comparison symbol $\sim$ used in
  expressions like $x \sim \alpha$ and $\alpha \sim \beta$ belongs to the
  extended set $\{ = , < , \leq , > , \geq, \cleq, \cgeq \}$. For any constant
  $a \in \N^+$, notation $z \cleq z'$ means $z \equiv z' \bmod a$ and $z \leq
  z'$. Equivalently, this means that there exists $y \in \N$ such that $z + ay
  = z'$. Notation $z \cgeq z'$ means $z \equiv z' \bmod a$ and $z \geq z'$.

\noindent
Any $x \sim \alpha$ is called an \emph{$x$-atom}, any $\alpha \sim \beta$ is
called a \emph{$\theta$-atom}. An \emph{\xconj} is any conjunction of
$x$-atoms, and a \emph{\tconj} is any conjunction of $\theta$-atoms. We denote
by ${\sf B}_{x,\Theta}$ the set of \emph{boolean combinations} of $x$-atoms
and $\theta$-atoms. A \emph{guard} is any element of ${\sf B}_{x,\Theta}$.
Thus the set $\sf G$ of Notation \ref{not:guard} is now equal to the set
${\sf B}_{x,\Theta}$.
\end{notation}

From now on, it is supposed that the guards and the invariants appearing in
parametric timed automata belong to the generalized set ${\sf G} = {\sf
  B}_{x,\Theta}$. It should be noted that the extension of $\sim$ to $\{ = , <
, \leq , > , \geq, \cleq, \cgeq \}$ is only valid inside automata, and
\emph{not} inside \logic\ formulae. We shortly call \emph{automaton} any
parametric timed automaton $\sf A$.

The next lemma states that any ${\sf B}_{x,\Theta}$ formula is a Presburger
formula. It also states that this formula can be rewritten in a particular
form that will be useful later.

\begin{lemma}   \label{lem:normal_form}
  Any ${\sf B}_{x,\Theta}$ formula is a Presburger formula. It can be
  rewritten as a disjunction of conjunctions of $x$-atoms and $\theta$-atoms
  with $\sim$ limited to $\{ = , \leq , \geq , \cleq, \cgeq \}$.
\end{lemma}

\proof
  Operators $\cleq$ and $\cgeq$ are easily rewritten in Presburger arithmetic.
  Even if linear terms $\alpha, \beta, \ldots$ contain constants in $\Z$, any
  $x \sim \alpha$ and $\alpha \sim \beta$ can also be rewritten in Presburger
  arithmetic. This shows that any ${\sf B}_{x,\Theta}$ formula is a
  Presburger formula.
  
  To rewrite a ${\sf B}_{x,\Theta}$ formula as described in the lemma, it is
  first put into disjunctive normal form. Second negation is suppressed in any
  $\neg (z \sim z')$ as follows. This is done easily for $\sim \; \in
  \{<,\leq,>,\geq\}$. Negation $\neg (z = z')$ is replaced by $z < z' \vee z >
  z'$. Negation $\neg(z \cleq z')$ is equivalent to $(z > z') \vee
  (\bigvee_{0<b<a} z+b \cleq z')$.  Similarly for $\neg(z \cgeq z')$. Third
  all inequalities $z < z'$ and $z > z'$ are replaced respectively by $z \leq
  z'-1$ and $z \geq z'+1$. Finally this formula is put into disjunctive normal
  form.
\qed

 Let us now state our main result.

\begin{thm} \label{thm:main}
  Let $\sf A$ be an automaton and $q$ be a state of $\sf A$. Let $\varphi$ be
  a {\rm QF-F-}\logic. Then there exists a ${\sf B}_{x,\Theta}$ formula
  $\Delta_{q,\varphi}(x,\Theta)$ with $x$ and all $\theta \in \Theta$ as free
  variables such that
  $$(q,x_0) \models_v \varphi \quad\mbox{iff}\quad
  \Delta_{q,\varphi}(x_0,v({\Theta})) \mbox{ is {\sc true}}$$
  for any
  valuation $v$ on $\Theta$ and any clock value $x_0$. The construction of
  formula $\Delta_{q,\varphi}$ is effective.
\end{thm}

The proof of Theorem \ref{thm:main} is by induction on the way formula
$\varphi$ is constructed. Before detailing its proof, we roughly give the main
ideas.  First, suppose for instance that along a run $\rho = (q_i,x_i)_{i\geq
  0}$ of ${\sf A}^v$ showing that $(q_0,x_0) \models_v \varphi$, some
configuration, say $(q_j,x_j)$, needs to satisfy $(q_j,x_j) \models_v \psi$
with $\psi$ a subformula of $\varphi$. The automaton $\sf A$ is modified into
${\sf A}'$ such that the invariant ${\sf I}(q_j)$ is
\emph{augmented}\footnote{Such kind of invariant is allowed in Notation
  \ref{not:ext_guard}.} by the ${\sf B}_{x,\Theta}$ formula
$\Delta_{q_j,\psi}$ constructed by induction. Along the run $\rho$ seen in the
modified automaton ${\sf A}'$, the satisfaction relation $(q_j,x_j) \models_v
\psi$ holds automatically thanks to the augmented invariant of $q_j$. Second,
what we also need is a ${\sf B}_{x,\Theta}$ formula that expresses the
existence of an infinite run starting at a given configuration (for operator
$\EG$ for instance) and another one that expresses the existence of a finite
run $\rho$ starting and ending at given configurations such that $\Time_{\rho}
\sim v(\alpha)$ (for operator $\EU_{\sim \alpha}$ for instance). This is
possible by the next two propositions. Their proofs are postponed till Section
\ref{sec:durations}.

\begin{proposition}     \label{prop:progress}
  Let ${\sf A}$ be an automaton and $q$ be a state. Then there exists a
  ${\sf B}_{x,\Theta}$ formula $\Prog_q(x,\Theta)$ such that for any
  valuation $v$ and any clock value $x_0$,
$$\Prog_q(x_0,v(\Theta)) \mbox{ is {\sc true}}$$
iff there exists an infinite
run in ${\sf A}^v$ starting with $(q,x_0)$. The construction of
$\Prog_q(x,\Theta)$ is effective.
\end{proposition}

\begin{proposition} \label{prop:lambda}
  Let ${\sf A}$ be an automaton and $q, q'$ be two states. Let $\sim \; \in
  \{<,\leq,>,\geq\}$ and $\alpha$ be a linear term. Then there exists a ${\sf
    B}_{x,\Theta}$ formula $\llambda_{q,q'}^{\sim\alpha}(x,\Theta)$ such that
  for any valuation $v$ and any clock value $x_0$,
$$\llambda_{q,q'}^{\sim\alpha}(x_0,v(\Theta)) \mbox{ is {\sc true}}$$
iff
there exists a finite run $\rho = (q,x_0) \pAth (q',\cdot)$ in ${\sf A}^v$
with $\Time_{\rho} \sim v(\alpha)$. The construction of
$\llambda_{q,q'}^{\sim\alpha}(x,\Theta)$ is effective.
\end{proposition}

For the proof of Theorem \ref{thm:main}, instead of the grammar given in
Definition \ref{def:fragment}, we prefer to work with the grammar
\begin{eqnarray*}
\varphi &::=& \sigma ~|~ \alpha \sim \beta ~|~ \neg \varphi ~|~ \varphi \vee
  \varphi ~|~ \EX \varphi \\
 & &~|~\varphi \EU_{< \alpha} \varphi ~|~ \varphi
  \EU_{\leq \alpha} \varphi ~|~\varphi \EU_{> \alpha} \varphi ~|~\varphi
  \EU_{\geq \alpha} \varphi \\
 & &~|~ \EG_{ < \alpha} \varphi ~|~ \EG \varphi
\end{eqnarray*}
This grammar is equivalent because formula $\varphi \AU_{\sim\alpha} \psi$
with $\sim \in \{<, \leq\}$ can be replaced by $\neg[(\EG_{\sim\alpha} \neg
\psi) \vee (\neg \psi \EU_{\sim\alpha} (\neg \varphi \wedge \neg \psi)) ]$,
formula $\varphi \AU \psi$ by $\neg[(\EG \neg \psi) \vee (\neg \psi \EU (\neg
\varphi \wedge \neg \psi)) ]$, and formula $\EG_{\leq \alpha} \varphi$ by
$\EG_{< \alpha+1} \varphi$.

It is not difficult to check that the semantics of the new operator $\EG_{ <
  \alpha} \varphi$ is given by

\medskip
\begin{minipage}{11.5cm}
  $(q,x) \models_v \EG_{ < \alpha} \varphi$ iff there exists a run $\rho =
  (q_i,x_i)_{i \geq 0}$ of ${\sf A}^v$ with $(q,x) = (q_0,x_0)$, there exists
  $j \geq 0$ such that $D_{\rho}(q_j,x_j) \geq v(\alpha)$ and $(q_i,x_i)
  \models_v \varphi$ for all $i < j$.
\end{minipage}

\bigskip

\proof (of Theorem \ref{thm:main}). The proof is by induction on
  $\varphi$.
\begin{enumerate}[$\bullet$]
\item If $\varphi = \sigma$, then $(q,x_0) \models_v \varphi$ iff there exists
  an infinite run starting with $(q,x_0)$ and $\sigma \in {\sf L}(q)$.
  Therefore
 $$\begin{array}{llll}
        \Delta_{q,\varphi}(x,\Theta) 
            &=& \false      & \mbox{if } \sigma \notin {\sf L}(q) \\
            &=& \Prog_{q}(x,\Theta) & \mbox{otherwise}.
\end{array}$$
\item Similarly, if $\varphi = \alpha \sim \beta$ with $\sim \; \in
\{=,<,\leq,>,\geq\}$, then
$$\begin{array}{llll}
        \Delta_{q,\varphi}(x,\Theta) 
            &=& (\alpha \sim \beta) \wedge \Prog_{q}(x,\Theta).
\end{array}$$
\item If $\varphi = \psi \vee \phi$, then $\Delta_{q,\varphi} =
  \Delta_{q,\psi} \vee \Delta_{q,\phi}$.
\item If $\varphi = \neg \psi$, then $\Delta_{q,\varphi} = \neg
  \Delta_{q,\psi}$.
\item Let us treat $\varphi = \EX \psi$. Recall that $(q,x_0) \models_v \EX
  \psi$ iff there exists a transition $(q,x_0) \stackrel{\tau}{\rightarrow}
  (q',x'_0)$ such that $(q',x'_0) \models_v \psi$ and $(q',x'_0)$ is the first
  configuration of an infinite run $\rho'$. Let $(q,\tau,g,r,q')$ be the edge
  of $E$ that has lead to the transition $(q,x_0) \stackrel{\tau}{\rightarrow}
  (q',x'_0)$. Then (see Definition \ref{def:config}), $x'_0 = 0$ if $r =
  \{x\}$, and $x'_0 = x_0 + \tau$ if $r = \emptyset$. By induction hypothesis,
  $\Delta_{q',\psi}$ has been constructed such that
  $\Delta_{q',\psi}(x'_0,v(\Theta))$ is {\sc true} iff $(q',x'_0) \models_v
  \psi$. The automaton $\sf A$ is modified into an automaton $\overline{\sf
    A}$ as follows. A copy\footnote{The copy $\overline{q}'$ of $q'$ is needed
    to focus on the first configuration $(q',x'_0)$ of $\rho'$.}
  $\overline{q}'$ of $q'$ is added to $Q$ such that ${\sf L}(\overline{q}') =
  {\sf L}(q')$, ${\sf I}(\overline{q}') = {\sf I}(q') \wedge
  \Delta_{q',\psi}(x,\Theta)$. A copy $(\overline{q}',\tau',g',r',p)$ is also
  added for each edge $(q',\tau',r',g',p)$ leaving $q'$. By Proposition
  \ref{prop:progress} applied to $\overline{\sf A}$ and $\overline{q}'$, we
  get a ${\sf B}_{x,\theta}$ formula $\Prog_{\overline{q}'}$ such that
  $\Prog_{\overline{q}'}(x'_0,v(\Theta))$ is {\sc true} iff there exists an
  infinite run in $\overline{\sf A}^v$ starting with $(\overline{q}',x'_0)$.
  By construction of $\overline{q}'$, equivalently there exists an infinite
  run in ${\sf A}^v$ starting with $(q',x'_0)$ and such that $(q',x'_0)
  \models_v \psi$. Hence, the expected formula $\Delta_{q,\varphi}(x,\Theta)$
  is equal to
$$\begin{array}{llll} 
  \Delta_{q,\varphi}(x,\Theta) &=&&     
        \bigvee_{(q,\tau,g,\{x\},q') \in E} \;
        ({\sf I}(q) \wedge \Prog_{\overline{q}'}(0,\Theta)) \\
        && \vee &  \bigvee_{(q,\tau,g,\emptyset,q') \in E} \;
  ({\sf I}(q) \wedge \Prog_{\overline{q}'}(x+\tau,\Theta)).
\end{array}$$
\item The construction of formula $\Delta_{q,\varphi}$ for $\varphi = \EG
  \psi$ is in the same vein as the previous one. Recall that $(q,x_0)
  \models_v \varphi$ iff there is an infinite run in ${\sf A}^v$ with first
  configuration $(q,x_0)$ such that all its configurations satisfy $\psi$. The
  automaton $\sf A$ is here modified into $\overline{\sf A}$ as follows. For
  any state $p \in Q$, ${\sf I}(p)$ is replaced by ${\sf I}(p) \wedge
  \Delta_{p,\psi}(x,\Theta)$. By Proposition \ref{prop:progress} applied to
  $\overline{\sf A}$, we get a formula $\Prog_{q}$ such that
  $\Prog_{q}(x_0,v(\Theta))$ is {\sc true} iff there exists an infinite run in
  ${\sf A}^v$ starting with $(q,x_0)$ and such that all its configurations
  satisfy $\psi$. Therefore formula $\Delta_{q,\varphi}(x,\Theta)$ is equal to
$$\Prog_{q}(x,\Theta).$$
\item Let us turn to formula $\varphi = \psi \EU_{\sim \alpha} \phi$ with $\sim
  \in \{<,\leq,>,\geq\}$. We have $(q,x_0) \models_v \varphi$ iff either (1)
  $(q,x_0) \models_v \phi$, $0 \sim v(\alpha)$ and $(q,x_0)$ is the first
  configuration of an infinite run, or (2) there exists a finite run $\rho =
  (q,x_0) \pAth (q',x'_0)$ such that $\Time_{\rho} \sim v(\alpha)$, $\psi$ is
  satisfied at every configuration of $\rho$ distinct from $(q',x'_0)$, $\phi$
  is satisfied at $(q',x'_0)$ and $(q',x'_0)$ is the first configuration of an
  infinite run. For any state $p \in Q$, formulae $\Delta_{p,\psi}$ and
  $\Delta_{p,\phi}$ have been constructed by induction hypothesis. So, in case
  (1), with the same construction of $\overline{\sf A}$ as done before for
  operator $\EX$ (with $q$, $\phi$ instead of $q'$, $\psi$), we have the next
  formula
$$(0 \sim \alpha) \wedge \Prog_{\overline{q}}(x,\Theta).$$
Case (2) is more
involved. The automaton $\sf A$ is first modified into $\overline{\sf A}$ as
for operator $\EX $ (with $q', \phi$ instead of $q', \psi$) to get formula
$\Prog_{\overline{q}'}$ such that $\Prog_{\overline{q}'}(x'_0,v(\Theta))$ is
{\sc true} iff there exists an infinite run in ${\sf A}^v$ starting with
$(q',x'_0)$ and such that $(q',x'_0) \models_v \phi$. The automaton $\sf A$
is then modified in another automaton $\underline{\sf A}$ in the following
way. A copy $\underline{q}'$ of $q'$ is added to $Q$ as well as a copy of each
edge of $E$ entering $q'$ as entering $\underline{q}'$; we define ${\sf
  L}(\underline{q}') = {\sf L}(q')$ and ${\sf I}(\underline{q}') = {\sf
  I}(q') \wedge \Prog_{\overline{q}'}(x,\Theta)$.\footnote{The copy
  $\underline{q}'$ of $q'$ is needed to focus on the last configuration
  $(q',x'_0)$ of $\rho$; the augmented invariant is needed to express that
  $\phi$ is satisfied at $(q',x'_0)$ and $(q',x'_0)$ is the first
  configuration of an infinite run.} For any state $p$ of $Q$, ${\sf I}(p)$
is replaced by ${\sf I}(p) \wedge \Delta_{p,\psi}(x,\Theta)$. Thanks to
Proposition \ref{prop:lambda} applied to $\underline{\sf A}$, we obtain a
formula $\llambda_{q,\underline{q}'}^{\sim\alpha}(x,\Theta)$ expressing the
following: $\llambda_{q,\underline{q}'}^{\sim\alpha}(x_0,v(\Theta))$ is {\sc
  true} iff there exists in $\underline{\sf A}^v$ a finite run
$\underline{\rho} = (q,x_0) \pAth (\underline{q}',x'_0)$ with
$\Time_{\underline{\rho}} \sim v(\alpha)$. Equivalently there exists in ${\sf
  A}^v$ a finite run $\rho = (q,x_0) \pAth (q',x'_0)$ with $\Time_{\rho} \sim
v(\alpha)$ such that $\psi$ is satisfied at every configuration of $\rho$
distinct from $(q',x'_0)$, $\phi$ is satisfied at $(q',x'_0)$ and $(q',x'_0)$
is the first configuration of an infinite run. For case (2), the expected
formula is thus the disjunction
$$\bigvee_{q' \in Q} \llambda_{q,\underline{q}'}^{\sim\alpha}(x,\Theta).$$
Therefore, putting together cases (1) and (2), formula $\Delta_{q,\varphi}$ is
the disjunction
$$\left((0 \sim \alpha) \wedge \Prog_{\overline{q}}(x,\Theta)\right)
\quad\vee\quad \bigvee_{q' \in Q}
\llambda_{q,\underline{q}'}^{\sim\alpha}(x,\Theta).$$
\item Finally, let $\varphi$ be $\EG_{< \alpha} \psi$. Then $(q,x_0) \models_v
  \varphi$ iff there exists a finite run $\rho = (q,x_0) \pAth (q',x')$ such
  that $\Time_{\rho} \geq v(\alpha)$, $(p,x) \models_v \psi$ for each
  configuration $(p,x)$ of $\rho$ distinct from $(q',x')$ and $(q',x')$ is the
  first configuration of an infinite run. As done just before in case (2),
  $\sf A$ is modified into $\underline{\sf A}$ except that we use
  $\Prog_{q'}$ instead of $\Prog_{\overline{q}'}$ in the definition of ${\sf
    I}(\underline{q}')$. By Proposition \ref{prop:lambda}, formula
  $\Delta_{q,\varphi}$ is equal to
          $$\bigvee_{q' \in Q}
          \llambda_{q,\underline{q}'}^{\geq\alpha}(x,\Theta).$$
\end{enumerate}
The proof is completed since all the proposed formulae belong to ${\sf
  B}_{x,\Theta}$ and their construction is effective.
\qed

Solutions to the model-checking problem and the parameter synthesis problem
are obtained as a corollary of Theorem \ref{thm:main}.

\begin{cor}   \label{cor:1}
The model-checking problem for {\rm F-}\logic\ is decidable. 
\end{cor}

\proof
  Let $Q \Theta \; \varphi$ be a F-\logic\ formula $f$ with no free parameters.
  By Theorem~\ref{thm:main}, 
  $$(q,x_0) \models f \quad\mbox{ iff }\quad Q \Theta \;
  \Delta_{q,\varphi}(x_0,\Theta) \mbox{ is {\sc true}}.$$
  By
  Lemma~\ref{lem:normal_form}, formula $Q \Theta \;
  \Delta_{q,\varphi}(x_0,\Theta)$ is a Presburger formula.  As Presburger
  arithmetic has a decidable theory and $Q \Theta \;
  \Delta_{q,\varphi}(x_0,\Theta)$ is a Presburger sentence, the model-checking
  problem is decidable.
\qed

The next corollary is straightforward. It states that the parameter synthesis
problem is solvable.

\begin{cor}   \label{cor:2}
  Let $\sf A$ be an automaton and $(q,x_0)$ a configuration of $\sf A$. Let
  $\{\theta_1, \ldots, \theta_k \}\subseteq\Theta$ with $k \geq 0$ and let $f
  = Q_1\theta_1 \; \cdots \; Q_k \theta_k \; \varphi$ be a {\rm F-}\logic\ 
  formula. Then the Presburger formula $Q_1\theta_1 \; \cdots \; Q_k \theta_k
  \; \Delta_{q,\varphi}(x_0,\Theta)$ with free variables in $\Theta_f$ is an
  effective characterization of the set of valuations $v$ on $\Theta_f$ such
  that $(q,x_0) \models_v f$.  \hfill$\Box$
\end{cor}

Corollary~\ref{cor:2} has important consequences that we want to detail
now.  Let us denote by $V({\sf A},f,q,x_0)$ the set of valuations $v$ on
$\Theta_f$ such that $(q,x_0) \models_v f$. Let $\Theta_f$ be equal to
$\{\theta'_1, \ldots, \theta'_l\}$. Presburger arithmetic has an effective
quantifier elimination, by adding to the operations $+$ and $\leq$ all the
congruences $\equiv \bmod a$, $a \in \N^+$. It follows the characterization of
$V({\sf A},f,q,x_0)$ given above in Corollary~\ref{cor:2} by
$$Q_1\theta_1 \; \cdots
\; Q_k \theta_k \; \Delta_{q,\varphi}(x,\Theta)$$
can be effectively rewritten
without any quantifier. On the other hand, since Presburger arithmetic has a
decidable theory, any question formulated in this logic about $V({\sf
  A},f,q,x_0)$ is decidable. For instance, the question ``Is the set $V({\sf
  A},f,q,x_0)$ non empty'' is decidable as it is formulated in Presburger
arithmetic by 
$$\exists \theta'_1 \; \cdots \; \exists \theta'_l \; Q_1\theta_1
\; \cdots \; Q_k \theta_k \; \Delta_{q,\varphi}(x,\Theta).$$
The question
``Does the set $V({\sf A},f,q,x_0)$ contain all the valuations on
$\Theta_f$'' is also decidable as it can be formulated as
$$\forall \theta'_1
\; \cdots \; \forall \theta'_l \; Q_1\theta_1 \; \cdots \; Q_k \theta_k \;
\Delta_{q,\varphi}(x,\Theta).$$
The question ``Is the set $V({\sf A},f,q,x_0)$
finite'' is translated into
$$\exists z \forall \theta'_1 \; \cdots \; \forall \theta'_l \; Q_1\theta_1 \;
\cdots \; Q_k \theta_k \; ~ (\Delta_{q,\varphi}(x,\Theta) \;\Rightarrow\;
\wedge_i \theta'_i \leq z).$$
And so on.

\section{Durations}     \label{sec:durations}

The aim of this section is to prove Propositions \ref{prop:progress} and
\ref{prop:lambda}. This is achieved thanks to a precise description of the
possible durations of finite runs in an automaton. Several steps are necessary
for this purpose.

In the first subsection, we show that we can work with automata put in some
normal form. This normalization allows a simplified presentation of the proofs
of the next subsections.

In Subsections \ref{subsec:aut_transf} and \ref{subsec:durations_free}, we
restrict to \emph{reset-free} normalized automata, that is automata in which
there is no reset of the clock. For this family of automata, we study the runs
of the form $(i,x_0) \pAth (f,\cdot)$ such that $i \in I$, $f \in F$ with $I$,
$F$ being two fixed subsets of states, and $x_0$ is a fixed clock value. In
Subsection \ref{subsec:aut_transf}, a sequence of transformations is performed
on the automata such that the $x$-atoms used in the automata are limited to
equalities $x = \alpha$. These simplifications lead in Subsection
\ref{subsec:durations_free} to the description by a Presburger formula of the
durations $D_{\rho}$ of runs $\rho = (i,x_0) \pAth (f,\cdot)$, $i \in I$, $f
\in F$.

In the last subsection, we remove the reset-free restriction imposed to the
automata and we study in details the durations $D_{\rho}$ of runs $\rho =
(q,x_0) \pAth (q',\cdot)$ between two fixed states $q$ and $q'$. Any such run
$\rho$ can be decomposed into a sequence of runs $\rho_j$, $1 \leq j \leq k$,
according to the reset of the clock, that is the clock is reset at the
beginning and the end of $\rho_j$ but not inside of $\rho_j$. The duration
$D_{\rho}$ of $\rho$ is thus the sum of the durations $D_{\rho_j}$, $1 \leq j
\leq k$. Any $D_{\rho_j}$ falls into durations being studied in Section
\ref{subsec:durations_free}. Thanks to this description of any duration
$D_{\rho}$ in terms of durations in reset-free automata, we are finally able
to prove Propositions \ref{prop:progress} and \ref{prop:lambda}.

In Subsections \ref{subsec:normalized}, \ref{subsec:aut_transf} and
\ref{subsec:durations_free}, we are going to perform a sequence of
transformations on the automata ${\sf A}$ that will \emph{preserve} the set
of runs in ${\sf A}^v$ for any valuation $v$, in the following sense. During
a transformation, state $q$ will possibly be splitted into several copies
$\overline{q}_j$. Runs before and after the splitting can be supposed
identical\footnote{Such an identification of runs is already present in the
  proof of Theorem \ref{thm:main}.} up to a \emph{renaming} of any
$\overline{q}_j$ into $q$.

\subsection{Normalized Automata}        \label{subsec:normalized}

In this subsection, the automata are put in some normal form. The aim of this
normalization is a simplified presentation of the proofs in the rest of the
paper.

\begin{defi}      \label{def:normalized}
An automaton $\sf A$ is \emph{normalized} if
\begin{enumerate}[$\bullet$]
\item The guards labeling the edges and used in the invariants are limited to
  conjunctions of $x$-atoms and $\theta$-atoms with $\sim \; \in \{ = , \leq ,
  \geq , \cleq, \cgeq \}$,
\item for any state $q \in Q$, the edges $(p,\tau,g,r,q)$ entering $q$ are all
  labeled by the same $g$ and the same $r$ (however $\tau$ can vary).
\end{enumerate}  
\end{defi}

\begin{proposition}     \label{prop:normalized}
  Any automaton $\sf A$ can be effectively normalized such that the set of
  runs in ${\sf A}^v$ is preserved for any valuation $v$.
\end{proposition}

\proof
  Let $g \in {\sf B}_{x,\Theta}$ be a guard. By Lemma \ref{lem:normal_form},
  it can be rewritten as a disjunction of $k$ formulae $\delta_j$, $1 \leq j
  \leq k$, where each $\delta_j$ is a conjunction of $x$-atoms and
  $\theta$-atoms with $\sim \; \in \{ = , \leq , \geq , \cleq, \cgeq \}$. If
  $g$ labels the edge $(q,\tau,g,r,q')$ of $\sf A$, then we modify $\sf A$
  by splitting this edge into $k$ edges $(q,\tau,\delta_j,r,q')$, $1 \leq j
  \leq k$. If $g = {\sf I}(q)$ for some state $q$, we modify $\sf A$ by
  splitting $q$ into $k$ states $\overline{q}_j$, $1 \leq j \leq k$, such that
  ${\sf L}(\overline{q}_j) = {\sf L}(q)$, ${\sf I}(\overline{q}_j) =
  \delta_j$ and we accordingly split any edge that enters or leaves state $q$.
  The first condition of Definition~\ref{def:normalized} is therefore
  satisfied.

  For the second condition, the construction is similar. Suppose that there
  are several edges $(p,\tau,g,r,q)$ entering state $q$ with distinct couples
  $(g,r)$. Then $q$ is splitted into several copies (one copy for one couple
  $(g,r)$) and all the edges entering $q$ are redirected to each copy,
  according to the couples $(g,r)$. The copies of $q$ have the same ${\sf
    L}(q)$ and ${\sf I}(q)$ as $q$.
\qed

\subsection{Transformations of Reset-free Automata}   \label{subsec:aut_transf}

In all this subsection, we assume the next hypothesis.

\medskip\noindent {\bf Hypothesis ($*$)} We assume that ${\sf A} =
(Q,I,F,E,{\sf L},{\sf I})$ is a \emph{reset-free normalized} automaton with
a set $I \subseteq Q$ of \emph{initial} states and a set $F \subseteq Q$ of
\emph{final} states. We also assume such that $I \cap F = \emptyset$, no edge
enters $i \in I$ and no edge leaves $f \in F$.

\medskip\noindent {\bf Remark} As $\sf A$ is normalized and reset-free, given
a state $q$, all edges $(p,\tau,g,r,q)$ entering $q$ have the same guard $g$
and satisfy $r = \emptyset$. It follows that we can move guard $g$ from these
edges to the invariant ${\sf I}(q)$ of $q$. Indeed $g$ is simply erased from
all the edges entering $q$ and added as a conjunction to ${\sf I}(q)$. By
this construction, the set $E$ of edges of $\sf A$ can be rewritten as a
subset of $Q \times \{0,1\} \times Q$, instead of $Q \times \{0,1\} \times
{\sf G} \times 2^{\{x\}} \times Q$ (see Definitions~\ref{def:timed_aut}
and~\ref{def:config}).

\noindent
On the other hand, as $\sf A$ is normalized, the invariant ${\sf I}(q)$ of
any state $q$ is a conjunction of $x$-atoms and $\theta$-atoms. We can view
${\sf I}(q)$ as a \emph{set} of $x$-atoms and $\theta$-atoms (instead of a
conjunction) and we will often say that an $x$-atom or a $\theta$-atom
\emph{belongs} to $q$ (instead of ${\sf I}(q)$) or \emph{appears} in $q$.

\medskip Given a valuation $v$ and a clock value $x_0$, we denote by
$${\sf R}({\sf A}^v,x_0)$$
the set of runs of ${\sf A}^v$ of the form
$(i,x_0) \pAth (f,\cdot)$ for some $i \in I$ and $f \in F$. We are going to
perform a sequence of transformations on ${\sf A}$ that will preserve ${\sf
  R}({\sf A}^v,x_0)$. The aim of these transformations is to simplify the
form of the invariants used in the automaton. The invariant ${\sf I}(q)$ of
any state $q \in Q \setminus (I \cup F)$ will be a conjunction of at most one
$x$-atom (of the form $x = \alpha$) and one \tconj. This simplification will
be possible mainly because the automaton is reset-free (see Proposition
\ref{prop:egalite}).

\begin{defi}      \label{def:simplified}
A reset-free normalized automaton ${\sf A}$ is \emph{simplified} if 
\begin{enumerate}[$\bullet$]
\item for all $q \in Q$, the invariant ${\sf I}(q)$ is equal to 
$${\sf I}_x(q) \wedge {\sf I}_{\theta}(q)$$ 
such that ${\sf I}_x(q)$ is an
  \xconj\ and ${\sf I}_{\theta}(q)$ is a \tconj. Among the $x$-atoms $x \sim
  \alpha$ of ${\sf I}_x(q)$, at most one is an equality $x = \alpha$.
  Moreover, if $q \not\in I \cup F$, then ${\sf I}_x(q)$ contains no other
  $x$-atom $x \sim \beta$ with $\sim \; \in \{\leq , \geq, \cleq, \cgeq \}$,
  and if $q \in I$ (resp. $q \in F$), then the other $x$-atoms of ${\sf
    I}_x(q)$ are of the form $x \geq \beta$ (resp. $x \leq \beta$).
\item for any run $\rho \in {\sf R}({\sf A}^v,x_0)$, for any $x$-atom $x =
  \alpha$, there exists at most one configuration $(q',x')$ of $\rho$ such
  that ${\sf I}_x(q')$ contains $x = \alpha$.
\end{enumerate}
\end{defi}

\noindent This definition is illustrated by the next very simple example.

\begin{figure}
        \begin{center}
                \includegraphics[scale=.45]{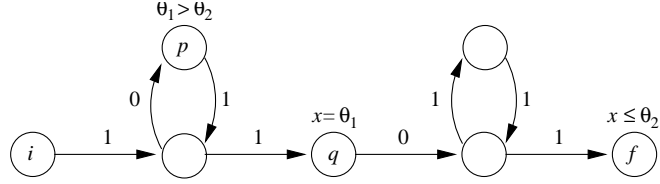}
        \end{center}
        \caption{A reset-free normalized automaton which is simplified}
        \label{fIg:ex}
\end{figure}

\medskip\noindent {\bf Example} Consider the simplified automaton $\sf A$ of
Figure \ref{fIg:ex} with one initial state $i$ and one final state $f$.
The invariant of state $p$ has no component ${\sf I}_x(p)$ and its
$\theta$-conjunction ${\sf I}_{\theta}(p)$ is limited to the $\theta$-atom
$\theta_1 > \theta_2$. The other states of the automaton has no
$\theta$-conjunction. They can have at most one $x$-atom which is an equality,
like state $q$ containing the equality $x = \theta_1$. The initial state $i$
can have $x$-atoms of the form $x \geq \alpha$ but it has no such $x$-atom
in this example. The final state $f$ has the $x$-atom $x \leq \theta_2$.

\begin{proposition}     \label{prop:egalite}
  Any reset-free normalized automaton $\sf A$ can be effectively simplified
  such that the set ${\sf R}({\sf A}^v,x_0)$ is preserved for any valuation
  $v$ and any clock value $x_0$.
\end{proposition}

\proof
  The proof of Proposition \ref{prop:egalite} needs several steps. The
  transformations described in the proof are based on standard constructions
  of automata theory. Each of them will preserve ${\sf R}({\sf A}^v,x_0)$
  for any valuation $v$ and any clock value $x_0$. After each transformation,
  the resulting automaton will be again denoted by $\sf A$.

  In the first step, we are going to suppress in each ${\sf I}_x(q)$, for $q
  \in Q$, all $x$-atoms of the form $x \cleq \alpha$.

\paragraph{First step} $x$-atoms $x \cleq \alpha$.

Let us show that any $x$-atom $x \cleq \alpha$ belonging to some state $q$ can
be suppressed at the cost of a new $x$-atom $x \leq \alpha$. The idea is the
following. If $\alpha \equiv b \bmod a$ for a certain $b \in
\{0,1,\ldots,a-1\}$\footnote{As $\alpha$ is a linear term over the parameters,
  the value $b$ such that $\alpha \equiv b \bmod a$ is not known whenever the
  parameter valuation $v$ is not fixed.},
then
$$x \cleq \alpha \quad \mbox{ iff } \quad x \equiv b \bmod a \mbox{ and } x
\leq \alpha.$$
The automaton is transformed in a way to compute modulo $a$.
New states are of the form $(q,c)$ with $q \in Q$ and $c \in \{0,\ldots,a-1\}$
expressing that $x \equiv c \bmod a$. Formally we construct ${\sf A}_b =
(Q',I',F',E',{\sf L}',{\sf I}')$ where $Q' = Q \times \{0,\ldots,a-1\}$, $I'
= I \times \{0,\ldots,a-1\}$, $F' = F \times \{0,\ldots,a-1\}$, ${\sf
  L}'(q,c) = {\sf L}(q)$ and $((q,c),\tau,(q',c')) \in E'$ iff $(q,\tau,q')
\in E$ and $c' \equiv c+\tau \bmod a$. Function ${\sf I'}$ is defined as
follows. For any $(q,c) \in Q'$, let ${\sf I}'(q,c) = {\sf I}(q)$. If
$(q,c)$ contains $x \cleq \alpha$, suppress this state if $c \neq b$, replace
$x \cleq \alpha$ by $x \leq \alpha$ if $c = b$. If $(q,c) \in I'$, add the
$x$-atom $x \cgeq c$ and the $\theta$-atom $\alpha \cgeq b$ to recall that
$\alpha \equiv b \bmod a$ and $x \equiv c \bmod a$ initially. As $\alpha$
depends on the parameter valuation, value $b$ such that $\alpha \equiv b \bmod
a$ is not known in advance. Therefore the final automaton is the disjoint
union of the automata ${\sf A}_b$, with $b \in \{0,\ldots,a-1\}$.

\medskip
The suppression of $x$-atoms $x \cgeq \alpha$ in each ${\sf I}_x(q)$ is
performed similarly. In the next step, we are going to suppress $x$-atoms $x
\geq \alpha$. This will be possible everywhere except inside states $q \in I$.

\paragraph{Second step} $x$-atoms $x \geq \alpha$.

Let us consider a fixed $x$-atom $x \geq \alpha$. Recall that the automaton is
reset-free. Along a run $\rho \in {\sf R}({\sf A}^v,x_0)$, as soon as $x
\geq \alpha$ is satisfied at some configuration of $\rho$, the next
occurrences of $x \geq \alpha$ are automatically satisfied and can be thus
suppressed. The automaton is transformed in a way to count occurrences of $x
\geq \alpha$ thanks to a counter $c$ equal to $0$ ($1$ or $2$ resp.) in case
of $0$ ($1$ or $2$ and more resp.) occurrence(s) of $x \geq \alpha$ is (are)
encountered.\footnote{Thus when the counter $c$ has value $2$, any
  incrementation $c+1$ lets it at value $2$.} Formally we construct ${\sf A}'
= (Q',I',F',E',{\sf L}',{\sf I}')$ where $Q' = Q \times \{0,1,2\}$, $F' = F
\times \{0,1,2\}$, ${\sf L}'(q,c) = {\sf L}(q)$ and ${\sf I}'(q,c) = {\sf
  I}(q)$ for all $q \in Q$ and $c \in \{0,1,2\}$. Sets $I'$ and $E'$ are
defined as follows. For any $q \in I$, state $(q,c)$ belongs to $I'$ with
$c=1$ if $x \geq \alpha$ belongs to $q$, and $c=0$ otherwise. For any
$(q,\tau,q') \in E$, edge $((q,c),\tau,(q',c'))$ belongs to $E'$ with $c' =
c+1$ if $q'$ contains $x \geq \alpha$, and $c' = c$ otherwise. Finally, we
suppress $x \geq \alpha$ in any state $(q,2)$ containing it.

Now, consider a run $\rho' \in {\sf R}({\sf A}'^v,x_0)$ equal to
$(q_i,c_i,x_i)_{0 \leq i \leq n}$ such that some state $(q_k,c_k)$ contains $x
\geq \alpha$. Necessarily, $c_k = 1$ and $c_i = 0$ for $0 \leq i < k$ by
construction of ${\sf A}'$. So $x$-atom $x \geq \alpha$ is satisfied at
configuration $(q_k,c_k,x_k)$ iff 
\begin{description}
\item[ ]\emph{(i)} either $x \geq \alpha$ is
satisfied at configuration $(q_0,c_0,x_0)$, 
\item[ ]\emph{(ii)} or $x = \alpha$ is
satisfied at some configuration $(q_i,c_i,x_i)$ of $\rho'$ such that $0 < i
\leq k$.
\end{description}
Therefore, $x$-atom $x \geq \alpha$ can be suppressed at the cost of
a new $x$-atom $x = \alpha$ (see \emph{(ii)}), except inside the initial state
$(q_0,c_0)$ (see \emph{(i)}). This can be achieved by modifying ${\sf A}'$
into an automaton ${\sf A}''$ thanks to a construction which is not difficult
but tedious, this will be not fully detailed. The automaton ${\sf A}''$ has
three parts :
\begin{enumerate}[$\bullet$] 
\item a first part of ${\sf A}''$ has to deal with paths of ${\sf A}'$ that
  only contain states $(q,c)$ with $c = 0$,
\item a second part has to deal with paths of ${\sf A}'$ starting with
  $(q,c)$ such that $q \in I$, $c = 1$, 
\item and a third part has to deal with paths of ${\sf A}'$ containing some
  state $(q,c)$ such that $q \not\in I$, $c = 1$; such paths are call
  \emph{special}.
\end{enumerate}
The first part of ${\sf A}''$ is obtained from ${\sf A}'$ by erasing all
states $(q,c)$ with $c = 1$. The second part is obtained from ${\sf A}'$ by
erasing all states $(q,c)$ such that $q \not\in I$, $c=1$ and all states
$(q,c)$ such that $q \in I$, $c=0$. We now discuss the third part of ${\sf
  A}''$. The special paths of ${\sf A}'$ must be modified into two kinds of
paths : either the $x$-atom $x \geq \alpha$ is added to the initial state of
the path (see \emph{(i)}), or the $x$-atom $x = \alpha$ is added to some
intermediate state of the path, which is situated between the initial state
(not included) and state $(q,c)$ (included) (see \emph{(ii)}). In both cases,
the $x$-atom $x \geq \alpha$ must be deleted from $(q,c)$. The third part of
${\sf A}''$, first case, is obtained from ${\sf A}'$ by adding the $x$-atom
$x \geq \alpha$ to any state $(q,c)$ such that $q \in I$, $c=0$ and by
deleting the $x$-atom $x \geq \alpha$ from any state $(q,c)$ such that $q
\not\in I$, $c=1$; it is also necessary to use a marker to verify that each
accepting path of ${\sf A}''$ corresponds to a special path of ${\sf A}'$.
The third part of ${\sf A}''$, second case, is obtained from ${\sf A}'$ as
follows : the $x$-atom $x \geq \alpha$ is deleted from any state $(q,c)$ such
that $q \not\in I$, $c=1$, all states $(q,c)$ with $q \not\in I$, $c=0$ are
duplicated (together with the edges entering and leaving $(q,c)$) such that
the $x$-atom $x = \alpha$ is added to one of the two copies of $(q,c)$; it is
also necessary to use a marker to verify that each accepting path of ${\sf
  A}''$ corresponds to a special path of ${\sf A}'$ and passes through
exactly one state containing the $x$-atom $x = \alpha$.

\medskip The suppression of $x$-atoms $x \leq \alpha$ can be performed in a
similar way. Note that here, as soon as the last (instead of the first)
occurrence of $x \leq \alpha$ is satisfied along a run $\rho \in {\sf
  R}({\sf A}^v,x_0)$, then the previous occurrences of $x \leq \alpha$ are
automatically satisfied. It follows that $x$-atoms $x \leq \alpha$ can be
suppressed everywhere except inside states $q \in F$.

\medskip At this point of the proof, for each state $q$, (1) if $q \not\in I
\cup F$, then the $x$-atoms contained in $q$ are of the form $x = \alpha$, (2)
if $q \in I$, then they are of the form $x = \alpha$ or $x \geq \alpha$, and
(3) if $q \in F$, then they are the form $x = \alpha$ or $x \leq \alpha$. It
remains to prove two facts about $x$-atoms which are equalities. First for all
$q \in Q$, among the $x$-atoms contained in $q$, at most one is an equality $x
= \alpha$. Second, for any run $\rho \in {\sf R}({\sf A}^v,x_0)$, for any
$x$-atom $x = \alpha$, there exists at most one configuration $(q',x')$ of
$\rho$ such that ${\sf I}_x(q')$ contains $x = \alpha$.

\paragraph{Third step} $x$-atoms $x = \alpha$.

The first fact can be easily proved. Suppose that ${\sf I}_x(q) =
\bigwedge_{\alpha \in \Alpha} (x = \alpha)$ for some set $\Alpha$ of linear
terms. Let $\alpha' \in \Alpha$. Then ${\sf I}_x(q)$ is equivalent to
$$(x = \alpha') \wedge \bigwedge_{\alpha \in \Alpha} (\alpha' = \alpha).$$
Thus ${\sf I}_x(q)$ can be replaced by $x = \alpha'$ and ${\sf
  I}_{\theta}(q)$ by ${\sf I}_{\theta}(q) \wedge \bigwedge_{\alpha \in
  \Alpha} (\alpha' = \alpha)$.

Let us prove the second fact. Let $\rho$ be a run in ${\sf R}({\sf
  A}^v,x_0)$. Assume that there are in $\rho$ several configurations
$(q_j,x_j)$, $1 \leq j \leq k$ such that $q_j$ contains a given $x$-atom $x =
\alpha$. It follows that time does not progress from $(q_1,x_1)$ to
$(q_k,x_k)$, that is, $x_j = x_1$ for all $j$. Only the first occurrence of $x
= \alpha$ at state $q_1$ is useful, the next ones can be forgotten. Therefore,
$\sf A$ is transformed in a way to count occurrences of $x = \alpha$ and to
remember any progress of time. As done before, a counter $c$ has value $0$
($1$ or $2$ resp.) in case of $0$ ($1$ or $2$ and more resp.) occurrences of
$x = \alpha$.  Moreover, values $1$ and $2$ are indexed by $+$ if time has
progressed since the first occurrence of $x = \alpha$. Formally we construct
${\sf A}' = (Q',I',F',E',{\sf L}',{\sf I}')$ where $Q' = Q \times
\{0,1,1_+,2,2_+\}$, $F' = F \times \{0,1,1_+,2,2_+\}$, ${\sf L}'(q,c) = {\sf
  L}(q)$ and ${\sf I}'(q,c) = {\sf I}(q)$ for all $q \in Q$ and $c \in
\{0,1,1_+,2,2_+\}$. For any $q \in I$, state $(q,c)$ belongs to $I'$ with
$c=1$ if $x = \alpha$ belongs to $q$, and $c=0$ otherwise. For any
$(q,\tau,q') \in E$, edge $((q,c),\tau,(q',c'))$ belongs to $E'$ where $c'$ is
computed according Table \ref{table:counter}.
\begin{table}[h]
\begin{minipage}{6cm}
\begin{center}
                $\begin{array}{|r||l|l|l|l|l|l|}
                \hline
                \tau \backslash c  & 0   & 1   & 1_+ & 2   & 2_+ \\
                \hline \hline
                       0 & 1   & 2   & 2_+ & 2   & 2_+ \\
                \hline 
                       1 & 1   & 2_+ & 2_+ & 2_+ & 2_+ \\
                \hline
                \end{array}$

                    if $q'$ contains $x = \alpha$

\end{center}
\end{minipage}
\begin{minipage}{6.0cm}
\begin{center}
    $\begin{array}{|r||l|l|l|l|l|l|}
                \hline
                \tau \backslash c  & 0   & 1   & 1_+ & 2   & 2_+ \\
                \hline \hline
                       0 & 0   & 1   & 1_+ & 2   & 2_+ \\
                \hline 
                       1 & 0   & 1_+ & 1_+ & 2_+ & 2_+ \\
                \hline
                \end{array}$
                otherwise

\end{center}
\end{minipage}
        \caption{Computation of $c'$}
        \label{table:counter}
\end{table}
Finally, for any state $(q,c)$ containing $x = \alpha$, we suppress this state
if $c = 2_+$, we suppress $x = \alpha$ from this state if $c = 2$. Indeed
recall that counter $2$ indicates that it is at least the second occurrence of
$x = \alpha$, and the presence of index $+$ means a progress of time since the
first occurrence of $x = \alpha$.
\qed

\subsection{Durations in Reset-free Automata}   \label{subsec:durations_free}

In this subsection, we again make Hypothesis ($*$). By
Proposition~\ref{prop:egalite}, we know that the reset-free normalized
automaton $\sf A$ can be supposed simplified. Thanks to this property of
$\sf A$, we are going to construct a Presburger formula describing all the
possible durations of runs in ${\sf R}({\sf A}^v,x_0)$ in terms of the
parameters. We need the next notation.

\begin{notation}
  Let $t$ be a variable used to denote a duration and $x$ be a variable for a
  clock value. We call \emph{$t$-atom} any $t \sim \alpha$ or $t \sim \alpha -
  x$, with $\alpha$ a linear term. A $t$-atom is of \emph{first type} if it is
  of the form
$$\begin{array}{l}
        t =     \alpha, \\
        t \cgeq \alpha, \\
        t =     \alpha - x,\\
        t \cgeq \alpha - x.
\end{array}$$
It is of \emph{second type} if it is of the form
$$\begin{array}{l}
t \leq \alpha - x.
\end{array}$$

\noindent 
A \emph{$t$-conjunction} is a conjunction of $t$-atoms of second type.
\end{notation}

\begin{proposition} \label{prop:lambda_reset_free}
  Let ${\sf A}$ be a reset-free normalized automaton. There exists a
  Presburger formula $\lambda(t,x,\Theta)$ such that for any valuation $v$ and
  any clock value $x_0$, there exists a run in ${\sf R}({\sf A}^v,x_0)$ with
  duration $t_0$ iff
$$\lambda(t_0,x_0,v(\Theta)) \mbox{ is {\sc true}}.$$
This formula is a disjunction of formulae of the form
$$\lambda_t \wedge \lambda_{\leq} \wedge \lambda_x \wedge \lambda_{\theta},$$
where $\lambda_t$ is a first type $t$-atom , $\lambda_{\leq}$ is a
$t$-conjunction, $\lambda_x$ is an \xconj\ and $\lambda_{\theta}$ is a \tconj.
Its construction is effective.
\end{proposition}

Let us explain this proposition on the next example.

\medskip\noindent {\bf Example} Consider the simplified automaton $\sf A$ of
Figure \ref{fIg:ex}. We denote by $t_0$ the duration of any run $(i,x_0) \pAth
(f,\cdot)$ in ${\sf R}({\sf A}^v,x_0)$, where $v$ is a fixed parameter
valuation. Every run has to pass through state $q$ which contains the $x$-atom
$x = \theta_1$. Let us study the possible durations $t_1$ of runs $\rho_1 =
(i,x_0) \pAth (q,\cdot)$.  Each duration $t_1$ must be equal to $v(\theta_1) -
x_0$. For runs $\rho_1$ using the cycle, constraint $v(\theta_1) >
v(\theta_2)$ holds and $t_1$ has the form $m + 3$, $m \geq 0$. The unique run
$\rho_1$ not using the cycle is not constrained and its duration equals $t_1 =
2$. Now any duration $t_0$ can be decomposed as $t_0 = t_1 + 2n + 1 =
v(\theta_1) - x_0 + 2n + 1$, $n \geq 0$. Due to the $x$-atom $x \leq \theta_2$
of state $f$, we get another constraint $x_0 + t_0 \leq v(\theta_2)$. In
summary, we have

$$\begin{array}{llllllllll}
  & [(v(\theta_1) - x_0 \conegeq 3 \;\;\wedge\;\; v(\theta_1) > v(\theta_2) ) &\vee&  v(\theta_1) - x_0 = 2] \\
  \wedge& [t_0 \ctwogeq v(\theta_1) - x_0 + 1] \\
  \wedge& [x_0 + t_0 \leq v(\theta_2)]
\end{array}$$

We get the next Presburger formula $\lambda(t,x,\Theta)$
$$\begin{array}{llllllllll}
  & [(x \coneleq \theta_1 - 3 \;\;\wedge\;\; \theta_1 > \theta_2 ) &\vee& x = \theta_1 - 2] \\
  \wedge& [t \ctwogeq \theta_1 + 1 - x] \\
  \wedge& [t \leq \theta_2 - x]
\end{array}$$
such that there exists a run in ${\sf R}({\sf A}^v,x_0)$ with duration $t_0$
iff $\lambda(t_0,x_0,v(\Theta))$ is {\sc true}. This formula is in the form of
Proposition \ref{prop:lambda_reset_free} when it is rewritten as a disjunction
of conjunctions of $t$-atoms, $x$-atoms and
$\theta$-atoms.\footnote{$\lambda_t$ is equal to $t \ctwogeq \theta_1 + 1 - x$
  and $\lambda_{\leq}$ is equal to $t \leq \theta_2 - x$.}

\medskip Thanks to the previous example, we can give some ideas of the proof
of Proposition \ref{prop:lambda_reset_free}. Except for the initial and final
states, the states of a simplified automaton contain at most one $x$-atom
which is of the form $x = \alpha$. The proof will be by induction on these
$x$-atoms. Given an $x$-atom $x = \alpha$ contained in some state $q$, any run
$\rho$ in ${\sf R}({\sf A}^v,x_0)$ passing through this state $q$ can be
decomposed as $(i,x_0) \pAth (q,x_1)$ and $(q,x_1) \pAth (f,x_2)$, for some
$i\in I$ and $f \in F$. Its duration $t_0$ can also be decomposed as $t_1 +
t_2$ with the constraint that the clock value $x_0 + t_1$ must satisfy $x =
\alpha$. It follows that $t_0 = v(\alpha) - x_0 + t_2$. The durations $t_1$
and $t_2$ and the related constraints will be computed by induction. When
there is no $x$-atom in the automaton (base case), only $\theta$-atoms can
appear in states. Runs will therefore be partitioned according to the set of
$\theta$-atoms that constrain them. Their durations will be described as fixed
values or arithmetic progressions.

\medskip
\proof (of Proposition \ref{prop:lambda_reset_free}). By
  Proposition~\ref{prop:egalite}, the reset-free normalized ${\sf A} =
  (Q,I,F,E,{\sf L},{\sf I})$ is assumed to be simplified.
  
\begin{enumerate}[(1)]
  \item We can suppose that $I$ is reduced to one initial state $i$ and
  $F$ to one final state $f$. At the end of the proof, it will remain to take
  a disjunction over $i \in I$ and $f \in F$ of the constructed formulae. From
  now on, we suppose that $I = \{i\}$ and $F = \{f\}$.

\item \emph{Assumption}. We make the assumption that $i$ contains no
$x$-atom and $f$ contains no $x$-atom $x \leq \alpha$. As $\sf A$ is
simplified, this means that for any state $q \in Q$, either ${\sf I}_x(q) =
\true$ or ${\sf I}_x(q)$ equals some $x = \alpha$. The proof is done by
induction on the $x$-atoms $x = \alpha$ that appear as ${\sf I}_x(q)$ with $q
\in Q$. The formula $\lambda(t,x,\Theta)$ that we will construct will have
\emph{no} $t$-conjunction, that is $\lambda(t,x,\Theta)$ will be a disjunction
of formulae of the form $\lambda_t \wedge \lambda_x \wedge \lambda_{\theta}$.

\medskip
\noindent \emph{Base case}.
Suppose that ${\sf I}_x(q) = \true$ for all $q \in Q$, that is ${\sf I}(q) =
{\sf I}_{\theta}(q)$. Durations of runs in ${\sf R}({\sf A}^v,x_0)$ are
thus independent on the clock values. They are simply equal to the number of
edges labeled by $\tau = 1$ along runs from $i$ to $f$. And to each of these
runs is associated a constraint which is the conjunction of the $\theta$-atoms
contained in the states of the run.

The proof is based on the classical Kleene theorem \cite{aut} using the
particular alphabet
$$B = \{ (\tau,\cc) \; | \; \tau \in \{0,1\}, \cc \in \{{\sf I}_{\theta}(q),q
\in Q\} \}.$$
To any edge $(q,\tau,q')$ of $\sf A$ corresponds the letter
$(\tau,{\sf I}_{\theta}(q'))$ of $B$. The concatenation $\cdot$ of two
letters $(\tau_1,\cc_1)$ and $(\tau_2,\cc_2)$ is defined as $(\tau_1 + \tau_2,
\cc_1 \wedge \cc_2)$. Thus a word over $B$ is equal to $(t,\cc)$ where $t$ is
a positive integer (a duration) and $\cc$ is a \tconj\ (a constraint on the
parameters). In particular, the empty word is equal to $(0,\true)$. The star
operation $^*$ is defined as usual and the plus\footnote{This notation should
  not be confused with the one used for the union operation.} operation $^+$
is defined by $L^+ = L^* \setminus \{(0,\true)\}$. We denote by
$\mbox{Rat}_B(\cdot,^+)$ the smallest family of languages containing $B$ and
closed under $\cdot$ and $^+$.  The elements of a set $L \in
\mbox{Rat}_B(\cdot,^+)$ have a simple form. The second components of these
elements are all identical because operation $\wedge$ is idempotent. The first
components constitute a set which is the union of a finite set and a finite
number of arithmetic progressions \cite{eilenberg}. In other words $L$ is
described by a disjunction of formulae of the form $\lambda_t \wedge
\lambda_{\theta}$ such that $\lambda_{\theta}$ equals a fixed \tconj\ $\cc$
and $\lambda_t$ equals either $t = \alpha$ or $t \cgeq \alpha$ with $\alpha
\in \N$.

Now by Kleene's theorem applied to $\sf A$, we get a rational language over
$B$ whose first components describe the durations of all runs of ${\sf
  R}({\sf A}^v,x_0)$ and the second components describe the related
constraints. It is not difficult to prove that this rational language can be
rewritten as a finite union of languages in $\mbox{Rat}_B(\cdot,^+)$. We thus
get the required formula $\lambda(t,x,\Theta)$ as a disjunction of formulae
$\lambda_t \wedge \lambda_{\theta}$ where $\lambda_t$ is a first-type $t$-atom
and $\lambda_{\theta}$ is a \tconj.

\medskip
\noindent \emph{General case}.
Now consider a particular $x$-atom $x = \alpha$. Let us denote by $P$ the set
of states $q$ such that ${\sf I}_x(q)$ is equal to $x = \alpha$. As $\sf A$
is simplified, any run $\rho$ of ${\sf R}({\sf A}^v,x_0)$ contains $0$ or
$1$ state of $P$ (see the second part of Definition \ref{def:simplified}). We
are going to prove that the expected formula $\lambda(t,x,\Theta)$ is equal to
$$\lambda^{Q \setminus P}(t,x,\Theta) \vee \bigvee_{p \in P}
\lambda^p(t,x,\Theta)$$
where $\lambda^{Q \setminus P}$ describes durations of
runs containing no state of $P$, and $\lambda^p$ describes durations of runs
containing one occurrence of the state $p$ of $P$.

All runs containing no state of $P$ constitute the set ${\sf R}({\sf
  A}'^v,x_0)$ of an automaton ${\sf A}'$ obtained from $\sf A$ by erasing
all states in $P$. As ${\sf A}'$ has one $x$-atom less, $\lambda^{Q\setminus
  P}(t,x,\Theta)$ can be constructed by induction hypothesis.

Let us now fix $p \in P$ and a run $\rho \in {\sf R}({\sf A}^v,x_0)$ that
contains it. This run is decomposed into a run $\rho_1 = (i,x_0) \pAth
(p,x_1)$ with duration $t_1$, and a run $\rho_2 = (p,x_1) \pAth (f,x_2)$ with
duration $t_2$. Duration $t_0$ of $\rho$ is equal to $t_1 + t_2$ such that
$x_1 = x_0 + t_1$, $x_2 = x_1 + t_2$ and $x_1$ satisfies $x = \alpha$.
Durations $t_1$ and $t_2$ can be computed by induction in the following way.

Let us begin with $t_1$. The automaton $\sf A$ is modified into ${\sf
  A}^{p,1}$ by erasing states of $P \setminus \{p\}$ and edges leaving $p$.
Invariant ${\sf I}_x(p)$ is replaced by $\true$. The new unique final state
is $p$. The new automaton has one $x$-atom less, so
$\lambda^{p,1}(t,x,\Theta)$ can be constructed by induction hypothesis such
that $\lambda^{p,1}(t_1, x_0, v(\Theta))$ is {\sc true}. Formula
$\lambda^{p,1}$ is a disjunction of formulae $\lambda^1_t \wedge \lambda^1_x
\wedge \lambda^1_{\theta}$ where $\lambda^1_t$ is a first type $t$-atom,
$\lambda^1_x$ is an \xconj\ and $\lambda^1_{\theta}$ is a \tconj. Suppose that
$\lambda^1_t$ is one among
        \begin{eqnarray}        \label{eq:1}
        t = \alpha_1, \quad t \cgeq \alpha_1, \quad t = \alpha_1 - x, \quad t \cgeq \alpha_1 - x.
        \end{eqnarray}
As $x_1$ satisfies $x = \alpha$ and $x_1 = x_0 + t_1$, then
        \begin{eqnarray}  \label{eq:2}
        x_1 = v(\alpha), \quad t_1 = v(\alpha) - x_0.
        \end{eqnarray}
So in (\ref{eq:1}), $t$ can be replaced by $\alpha - x$ and (\ref{eq:1}) becomes
$$\alpha - x = \alpha_1, \quad \alpha - x \cgeq \alpha_1, \quad \alpha =
\alpha_1, \quad \alpha \cgeq \alpha_1.$$
Thus $\lambda^1_t$ becomes an
$x$-atom or a $\theta$-atom. The modified formula $\lambda^1_t \wedge
\lambda^1_x \wedge \lambda^1_{\theta}$ is denoted by
   \begin{eqnarray}     \label{eq:1bis}
   \lambda'^1_x \wedge \lambda'^1_{\theta}.
   \end{eqnarray}

   Let us now describe $t_2$. We modify $\sf A$ into ${\sf A}^{p,2}$ by
   erasing states of $P \setminus \{p\}$ and edges entering $p$. Formula
   ${\sf I}_x(p)$ is replaced by $\true$. The new unique initial state is
   $p$. By induction hypothesis, $\lambda^{p,2}(t,x,\Theta)$ is constructed as
   a disjunction of formulae $\lambda^2_t \wedge \lambda^2_x \wedge
   \lambda^2_{\theta}$ where $\lambda^2_t$ is one among
        \begin{eqnarray}        \label{eq:3}
        t = \alpha_2, \quad t \cgeq \alpha_2, \quad t = \alpha_2 - x, \quad t \cgeq \alpha_2 - x.
        \end{eqnarray}
        
        Recall that $\lambda^{p,2}(t,x,\Theta)$ describes the duration $t_2$
        of runs $\rho_2 = (p,x_1) \pAth (f,x_2)$ for which $x_1$ satisfies $x
        = \alpha$. Thus in (\ref{eq:3}), $x$ can be replaced by $\alpha$ and
        (\ref{eq:3}) becomes
$$t = \alpha_2, \quad t \cgeq \alpha_2, \quad t = \alpha_2 - \alpha, \quad t
\cgeq \alpha_2 - \alpha.$$
This shows that $\lambda^2_t$ is now of the form
          \begin{eqnarray}      \label{eq:4}
          t = \beta \quad\mbox{or}\quad t \cgeq \beta.
          \end{eqnarray}
          Moreover $\lambda^2_x$ becomes a \tconj\ when $x$ is replaced by
          $\alpha$. The modified formula $\lambda^2_x \wedge
          \lambda^2_{\theta}$ is denoted by \begin{eqnarray} \label{eq:4bis}
            \lambda'^2_{\theta}.
   \end{eqnarray}

Finally, we can describe $t_0 = t_1 + t_2$. By (\ref{eq:2}) and (\ref{eq:4}),
it has the form 
\begin{eqnarray} \label{eq:5}
t_0 = v(\alpha) - x_0 + v(\beta) \quad \mbox{or}\quad t_0 \cgeq v(\alpha) -
x_0 + v(\beta).
\end{eqnarray}
                
Hence formula $\lambda^p(t,x,\Theta)$ for $t_0$ is a disjunction of formulae
$\lambda_t \wedge \lambda_x \wedge \lambda_{\theta}$ such that $\lambda_t$ has
the form (see (\ref{eq:5})) $t = \alpha - x + \beta$ or $t \cgeq \alpha - x +
\beta$ and $\lambda_x \wedge \lambda_{\theta}$ has the form (see
(\ref{eq:1bis} and (\ref{eq:4bis})) $\lambda'^1_x \wedge \lambda'^1_{\theta}
\wedge \lambda'^2_{\theta}$.

\item Under the assumption that $i$ contains no $x$-atoms and $f$
contains no $x$-atom $x \leq \alpha$, we have constructed a formula
$\lambda(t,x,\Theta)$ with no $t$-conjunction. So we have to take into account
the \xconj\ ${\sf I}_x(i)$ and the $x$-atoms $x \leq \alpha$ appearing in
$f$. Thus $x_0$ must satisfy ${\sf I}_x(i)$ and $x_0 + t_0$ must satisfy all
$x \leq \alpha$ in $f$. It follows that the final formula is equal to
\begin{eqnarray} \label{eq:x=0}
       \lambda(t,x,\Theta) \wedge {\sf I}_x(i)(x,\Theta) \wedge \bigwedge_{x \leq \alpha \in f} t \leq \alpha - x.
\end{eqnarray}
\end{enumerate}
\qed

\begin{remark} \label{rem:x=0}
  Suppose that ${\sf A}$ is an automaton such that ${\sf I}(i)$ equals $x =
  0$ for each initial state $i \in I$.  Then formula $\lambda(t,x,\Theta)$ of
  Proposition \ref{prop:lambda_reset_free} contains the $x$-atom $x = 0$ (see
  (\ref{eq:x=0})). Hence, if $\lambda(t_0,x_0,v(\Theta))$ is {\sc true}, then
  necessarily $x_0 = 0$, which can been interpreted as a reset of the clock.
  This remark will be used in the next subsection.
\end{remark}

\subsection{Durations in General}       \label{subsec:durations}

This subsection is devoted to the proofs of Propositions \ref{prop:progress}
and \ref{prop:lambda}. Here there is no longer the restriction on the
automaton given by Hypothesis ($*$): it is \emph{any} automaton as in
Definition \ref{def:timed_aut}. This automaton is supposed to be normalized by
Proposition \ref{prop:normalized}. Thus, given a state $q$, the edges
$(p,\tau,g,r,q)$ entering $q$ all have the same $r$. We call $q$ a
\emph{reset-state} in case $r = \{x\}$. The set of reset-states of $\sf A$ is
denoted by $Q_R$.

Let ${\sf A} = (Q,E,{\sf L},{\sf I})$ be an automaton. Let us fix two
states $q,q'$, a parameter valuation $v$, a clock value $x_0$. We denote by
$${\sf R}_{q,q'}({\sf A}^v,x_0)$$
the set of runs $\rho = (q,x_0) \pAth
(q',\cdot)$ in ${\sf A}^v$. Let us study this set.

A run $\rho$ in ${\sf R}_{q,q'}({\sf A}^v,x_0)$ possibly contains some
reset-states. It thus decomposes as a sequence of $k \geq 1$ runs $\rho_j$, $1
\leq j \leq k$, such that for any $j$, $\rho_j$ contains no reset-state,
except possibly for the first and the last configurations of $\rho_j$. The
duration $\Time_{\rho_j}$ of each $\rho_j$ can be computed thanks to
Proposition \ref{prop:lambda_reset_free}. For any $j$, $1 \leq j \leq k$, let
us denote by $\lambda^j(t,x,\Theta)$ the Presburger formula corresponding to
$D_{\rho_j}$ which is a disjunction of formulae $\lambda_t \wedge
\lambda_{\leq} \wedge \lambda_x \wedge \lambda_{\theta}$. So the total
duration $\Time_{\rho}$ is equal to the sum $\Sigma_{1 \leq j \leq k}
\Time_{\rho_j}$. We will see that the durations $D_{\rho}$ of runs $\rho \in
{\sf R}_{q,q'}({\sf A}^v,x_0)$ can be symbolically represented thanks to
rational expressions on an alphabet whose letters are the formulae $\lambda_t
\wedge \lambda_{\leq} \wedge \lambda_x \wedge \lambda_{\theta}$ that appear in
the $\lambda^j(t,x,\Theta)$'s. Thanks to this symbolic description and because
our logic is the fragment F-\logic, we will be able to
prove Propositions \ref{prop:progress} and \ref{prop:lambda}. It should be
noted that the durations $D_{\rho}$ of runs $\rho \in {\sf R}_{q,q'}({\sf
  A}^v,x_0)$ cannot be described by a Presburger formula as in Proposition
\ref{prop:lambda_reset_free}, otherwise the model-checking problem for \logic\ 
would be decidable (see Corollary \ref{cor:undecidable}).

Let us now explain in details all these ideas.

\medskip In a first step, we construct from $\sf A$ several reset-free
normalized automata as in Hypothesis ($*$). The construction is a standard one
in automata theory. Runs $\rho_j$ mentioned before will be runs in these
automata and their durations will be described thanks to Proposition
\ref{prop:lambda_reset_free}.

\paragraph{First construction} For each couple $(p, p')$ of states of $\sf A$
such that $p \in \{q\} \cup Q_R$ and $p' \in \{q'\} \cup Q_R$, we construct
from $\sf A$ the following reset-free automaton ${\sf A}_{p,p'} =
(Q',I',F',E',{\sf L}',{\sf I}')$. The set $Q'$ of states is $(Q \setminus
Q_R) \cup \{\overline{p}, \overline{p}'\}$ where $\overline{p}, \overline{p}'$
are copies of $p,p'$. The unique initial state is $\overline{p}$ and the
unique final state is $\overline{p}'$. Let ${\sf L}'(\overline{p}) = {\sf
  L}(p)$ and ${\sf L}'(\overline{p}') = {\sf L}(p')$. Let ${\sf
  I}'(\overline{p})$ be equal to ${\sf I}(p)$ if $p = q$ and to $({\sf I}(p)
\wedge x=0) $\footnote{The $x$-atom $x = 0$ imposes a reset of the clock at
  state $p$ (see Remark \ref{rem:x=0})} if $p \neq q$. Let ${\sf
  I}'(\overline{p}')$ be equal to ${\sf I}(p')$ if $p' \not\in Q_R$ and to
$({\sf I}(p') \wedge x=0) $\footnote{As ${\sf A}_{p,p'}$ must satisfy
  Hypothesis ($*$), no reset can appears on the edges} if $p' \in Q_R$. The
set $E'$ of edges is the union of $E$ restricted to $Q \setminus Q_R$ with the
next set of new edges$^{11}$
$$\begin{array}{lll}
  (\overline{p},\tau,g,r,p_1)             & \mbox{if } (p,\tau,g,r,p_1)  \in E \\
  (p_1,\tau,g,\emptyset,\overline{p}')            & \mbox{if } (p_1,\tau,g,r,p') \in E \\
  (\overline{p},\tau,g,\emptyset,\overline{p}') & \mbox{if } (p,\tau,g,r,p')
  \in E.
\end{array}$$
In this way, automaton ${\sf A}_{p,p'}$ satisfies Hypothesis ($*$).

Let $p \in \{q\} \cup Q_R$ and $p' \in \{q'\} \cup Q_R$. We define $x_1$ to be
equal to $x_0$ if $p = q$, and to $0$ if $p \neq q$. The runs of ${\sf
  R}({\sf A}_{p,p'}^v, x_1)$ are exactly the non-empty runs $(p,x_1) \pAth
(p',\cdot)$ of ${\sf A}^v$ that pass through no reset-state (except possibly
the first and the last states of the run). The durations of runs in ${\sf
  R}({\sf A}_{p,p'}^v, x_1)$ are described by formula
$\lambda^{p,p'}(t,x,\Theta)$ of Proposition \ref{prop:lambda_reset_free}. This
formula is a disjunction $\bigvee_{j} \lambda^{p,p',j}$ of formulae
\begin{eqnarray} \label{eq:decomp}
\lambda^{p,p',j} = \lambda_t^{p,p',j} \wedge \lambda_{\leq}^{p,p',j} \wedge \lambda_x^{p,p',j} \wedge \lambda_{\theta}^{p,p',j}.
\end{eqnarray}

For each couple $(p,p')$ and each $j$, we \emph{associate} a distinct letter
$b_{p,p',j}$ to each formula $\lambda^{p,p',j}$. The set of all these letters
is denoted by $B$. We say that letter $b_{p,p',j}$ is a \emph{reset-letter} if
$p$ is a reset-state. The set of reset-letters is denoted $B_R$.

\medskip In a second step, we construct another automaton from $\sf A$ in a
way to show how a run of ${\sf R}_{q,q'}({\sf A}^v,x_0)$ is decomposed into
a sequence of runs $\rho_j$ according to reset-states of $\sf A$. This
automaton will be a classical automaton \cite{aut}.

\paragraph{Second construction} We construct an automaton $\sf B$ over the
alphabet $B$ as follows. The set of states equals $Q_R \cup \{q,q'\}$ and the
set of edges equals $\{ (p,b,p') \;|\; b = b_{p,p',j} \mbox{ for some $j$}
\}$. The unique initial (resp. final) state is $q$ (resp. $q'$).

So, any run $\rho$ of ${\sf R}_{q,q'}({\sf A}^v,x_0)$ is map into a path in
$\sf B$ from $q$ to $q'$ which indicates how $\rho$ is decomposed according
to reset-states of $\sf A$. The duration of $\rho$ is symbolically
represented by the word that labels the corresponding path in $\sf B$. Hence
the set of durations of runs of ${\sf R}_{q,q'}({\sf A}^v,x_0)$ is
\emph{symbolically represented} by the rational subset accepted by $\sf B$.
We denote by
$$L_{q,q'}$$
this subset of $B^*$. Any word of $L_{q,q'}$ has \emph{at most
  one} letter that is non reset (the first letter of the word).

\medskip We now study in details rational expressions over the alphabet $B$
and in particular the rational expression defining $L_{q,q'}$.

\paragraph{Rational expressions}

Let $L^+$ be denoting $L^* \setminus \{\epsilon\}$ with $\epsilon$ denoting
the empty word and $\mbox{Rat}_B(\cdot,^+)$ be the smallest family closed under
$\cdot$ and $+$, and containing $B$. One can prove that any rational language
over $B$ can be effectively rewritten as a finite union of languages in
$\{\epsilon\} \cup \mbox{Rat}_B(\cdot,^+)$. Therefore
\begin{eqnarray}  \label{eq:a}
        L_{q,q'} = \bigcup_i L_i 
        \end{eqnarray}
with
$$
L_i = \{\epsilon\} \quad\mbox{or}\quad L_i = \{b_i\} \quad\mbox{or}\quad
L_i = b_i \cdot K_i $$
such that $b_i \in B, K_i
\in\mbox{Rat}_{B_R}(\cdot,^+)$. The set ${\sf R}_{q,q'}({\sf A}^v, x_0)$ is
decomposed into
\begin{eqnarray}  \label{eq:b}
{\sf R}_{q,q'}({\sf A}^v, x_0) = \bigcup_i {\sf R}_i
\end{eqnarray}
according to (\ref{eq:a}).

An non empty word of $L_{q,q'}$ is a sequence $b_1b_2 \cdots b_n \in B^+$. The
first letter $b_1$ describes runs from state $q$ to some reset-state $p_1$,
the clock value at $q$ is $x_0$. Each letter $b_i$, $i \geq 2$, is a
reset-letter. If $2 \leq i < n$, $b_i$ describes runs from reset-state
$p_{i-1}$ to reset-state $p_i$, the clock value at $p_{i-1}$ is $0$. If $i =
n$, $b_i$ describes runs from reset-state $p_{n-1}$ to state $q'$, the clock
value at $p_{n-1}$ is $0$. Let
\begin{eqnarray}  \label{eq:lambda}
\lambda^i_t \wedge \lambda^i_{\leq} \wedge \lambda^i_x \wedge \lambda^i_{\theta}
\end{eqnarray}
be the formula associated to each letter $b_i$, $i \geq 1$ (see
(\ref{eq:decomp})). Whenever $i \geq 2$, $\lambda^i_x$ contains the $x$-atom
$x = 0$ by Remark \ref{rem:x=0} and Definition of automaton ${\sf A}_{p,p'}$.
In this case, we prefer\footnote{The sequence $b_1b_2 \cdots b_n$ symbolically
  represents certain runs of ${\sf R}_{q,q'}({\sf A}^v, x_0)$. We are only
  interested in the initial clock value $x_0$ treated by formula $\lambda^i_x$
  of $b_1$.} to work with the equivalent formula
\begin{eqnarray}  \label{eq:kappa}
\kappa^i_t \wedge \kappa^i_{\leq} \wedge \kappa^i_{\theta}
\end{eqnarray}
such that $x$ has been replaced by $0$ in (\ref{eq:lambda}) (in particular,
$\lambda_x$ becomes a \tconj). In this formula $\kappa^i_t$ is a $t$-atom of
the form $t = \alpha$ or $t \cgeq \alpha$, $\kappa^i_{\leq}$ is a conjunction
of $t$-atoms of the form $t \leq \alpha$ and $\kappa^i_{\theta}$ is a \tconj.

The concatenation $b_1 \cdot b_2 \cdot \cdots b_n$ is interpreted as follows.
It is the \emph{sum} $t_1 + t_2 + \cdots + t_n$ of the durations $t_1, t_2,
\ldots, t_n$ respectively described by $\lambda^1_t, \kappa^2_t, \ldots,
\kappa^n_t$. It is the \emph{conjunction} of the related constraints
$$(\lambda^1_{\leq} \wedge \kappa^2_{\leq} \wedge \cdots \kappa^n_{\leq})
\wedge \lambda^1_x \wedge (\lambda^1_{\theta} \wedge \kappa^2_{\theta} \wedge
\cdots \kappa^n_{\theta}).$$
Formulae $\lambda^1_{\leq}, \kappa^2_{\leq},
\ldots \kappa^n_{\leq}$ impose upper bounds on $t_1, t_2, \ldots, t_n$. The
\xconj\ imposes constraints on the clock value $x_0$. The \tconj\ 
$(\lambda^1_{\theta} \wedge \kappa^2_{\theta} \wedge \cdots
\kappa^n_{\theta})$ impose constraints on the parameters.

\medskip In the next lemmas, we show that certain properties of runs in ${\sf
  R}_i$ can be expressed in Presburger arithmetics thanks to the symbolic
representation $L_i$ of ${\sf R}_i$ (see (\ref{eq:a}) and (\ref{eq:b})).
After these lemmas, we will be fully equipped to prove
Propositions~\ref{prop:progress} and~\ref{prop:lambda}. Note that Proposition
\ref{prop:lambda} can only be proved with $\sim$ limited to
$\{<,\leq,>,\geq\}$, otherwise the model-checking problem for \logic\ would be
decidable.

\begin{lemma}   \label{lem:nonempty}
  One can construct a ${\sf B}_{x,\Theta}$ formula $\mbox{\rm
    NonEmpty}_{L_i}(x,\theta)$ such that for any valuation $v$ and any clock
  value $x_0$, $\mbox{\rm NonEmpty}_{L_i}(x_0,v(\theta))$ is {\sc true} iff
  ${\sf R}_i$ is non empty.
\end{lemma}

\proof
  Runs of ${\sf R}_i$ have durations that are symbolically represented by the
  words of $L_i$. Let us construct formula $\mbox{NonEmpty}_{L_i}$ by
  induction on the rational expression defining $L_i$ (see (\ref{eq:a})). This
  formula will be equal to $\eta_x \wedge \eta_{\theta}$ with $\eta_x$ an
  \xconj\ imposing constraints on the clock and $\eta_{\theta}$ a \tconj\ 
  imposing constraints on the parameters.

Suppose $L_i = \{\epsilon\}$, then $\mbox{NonEmpty}_{L_i}(x,\Theta)$ equals $x
= 0$ is $q$ is a reset-state and ${\sf I}(q)(x,\Theta)$ otherwise. Indeed,
under these constraints, ${\sf R}_i$ is non empty since it contains the empty
run with the null duration. Suppose that $L_i = \{b_i\}$ with $b_i \in B$ and
associated formula $\lambda^i_t \wedge \lambda^i_{\leq} \wedge \lambda^i_x
\wedge \lambda^i_{\theta}$. Recall that $\lambda^i_t$ is one among the
$t$-atoms $t = \alpha$, $t = \alpha - x$, $t \cgeq \alpha$ or $t \cgeq \alpha
- x$ and that $\lambda^i_{\leq}$ is of the form $\bigwedge_{\beta} t \leq
\beta - x$. It follows that the non emptiness of ${\sf R}_i$ can be expressed
thanks to the minimum duration $t = \alpha$ ($t = \alpha - x$ resp.) of runs
in ${\sf R}_i$. Then
\begin{eqnarray}        \label{eq:nonempty}
\mbox{NonEmpty}_{L_i}(x,\Theta) &=& (\bigwedge_{\beta} \alpha \leq \beta - x) \wedge \lambda_x \wedge  \lambda_{\theta} \\
\nonumber
                               ( &=& (\bigwedge_{\beta} \alpha \leq \beta) \wedge \lambda_x \wedge \lambda_{\theta} \quad\mbox{ resp.})
\end{eqnarray}

Suppose now that $L_i = b_i \cdot K_i$ with $b_i \in B$ and $K_i \in
\mbox{Rat}_{B_R}(\cdot,^+)$. Let us first prove by induction on the rational
expression defining $K_i$ that $\mbox{NonEmpty}_{K_i}(\Theta)$ equals some
\tconj\ $\eta_{\theta}.$\footnote{There is no term $\eta_x$ since $K_i
  \subseteq B_R^+$, that is, $x = 0$ (see (\ref{eq:kappa})).} Let $K_i =
\{b_i\}$ with $b_i \in B_R$. We obtain a formula similar to
(\ref{eq:nonempty}) where $x$ is replaced by $0$ (see(\ref{eq:kappa})), so
$$\mbox{NonEmpty}_{K_i}(\Theta) = (\bigwedge_{\beta} \alpha \leq \beta) \wedge
\kappa_{\theta}.$$

\begin{sloppy}
Suppose that $K_i = K \cdot K'$ and formulae
$\mbox{NonEmpty}_{K}$, $\mbox{NonEmpty}_{K'}$ have been constructed by
induction. Then $\mbox{NonEmpty}_{K_i}(\Theta) = \mbox{NonEmpty}_{K}(\Theta)
\wedge \mbox{NonEmpty}_{K'}(\Theta)$ because the non emptiness of ${\sf R}_i$
requires the non emptiness of both $K$ and $K'$. If $K_i = K^+$, then
$\mbox{NonEmpty}_{K_i}(\Theta) = \mbox{NonEmpty}_{K}(\Theta)$ because
conjunction in an idempotent operation. Finally for $L_i = b_i \cdot K_i$, we
get $\mbox{NonEmpty}_{L_i}(x,\Theta) = \mbox{NonEmpty}_{\{b_i\}}(x,\Theta)
\wedge \eta_{\theta}$ where $\mbox{NonEmpty}_{\{b_i\}}(x,\Theta)$ is formula
(\ref{eq:nonempty}) and $\eta_{\theta}$ is the formula just constructed for
$K_i$.
\end{sloppy}
\qed

\begin{lemma}   \label{lem:nonnull}
  One can construct a ${\sf B}_{x,\Theta}$ formula $\mbox{\rm
    NonNull}_{L_i}(x,\theta)$ such that for any valuation $v$ and any clock
  value $x_0$, $\mbox{\rm NonNull}_{L_i}(x_0,v(\theta))$ is {\sc true} iff
  ${\sf R}_i$ contains a run with a non null duration.
\end{lemma}

\proof
  The proof is in the same vein as for Lemma \ref{lem:nonempty} with a similar
  form $\eta_x \wedge \eta_{\theta}$ for $\mbox{NonNull}_{L_i}(x,\theta)$.

If $L_i = \{\epsilon\}$, then clearly $\mbox{NonNull}_{L_i}(x,\theta) =
\false$. If $L_i = \{b_i\}$ with $b_i \in B$ and associated formula
$\lambda^i_t \wedge \lambda^i_{\leq} \wedge \lambda^i_x \wedge
\lambda^i_{\theta}$. Let us study as before formulae $\lambda^i_t$ and
$\lambda^i_{\leq}$, where $\lambda^i_{\leq} = \bigwedge_{\beta} (t \leq \beta
- x)$. If $\lambda^i_t$ equals $t = \alpha$, then $t$ is non null iff $\alpha
> 0$. Then $\mbox{NonNull}_{L_i}(x,\Theta)$ is the formula $(\alpha > 0)
\wedge (\bigwedge_{\beta} \alpha \leq \beta - x) \wedge \lambda^i_x \wedge
\lambda^i_{\theta}$. When $\lambda^i_t$ is $t = \alpha - x$, we have a similar
formula with $t$ non null if $\alpha - x > 0$. If $\lambda^i_t$ equals $t
\cgeq \alpha$, then a possible non null value for $t$ is either $\alpha$ if
$\alpha > 0$ or $a$ if $\alpha = 0$. We get formula
$\mbox{NonNull}_{L_i}(x,\Theta)$ equal to $((\alpha > 0 \wedge
\bigwedge_{\beta} (\alpha \leq \beta - x)) \vee (\alpha = 0 \wedge
\bigwedge_{\beta} (a \leq \beta - x))) \wedge \lambda^i_x \wedge
\lambda^i_{\theta}.$ A similar argument holds if $\lambda^i_t$ equals $t \cgeq
\alpha - x$.

Let $L_i = b_i \cdot K_i$, with $b_i \in B$ and $K_i \in
\mbox{Rat}_{B_R}(\cdot,+)$. Let us first construct formula
$\mbox{NonNull}_{K_i}(\Theta)$ by induction on $K_i$. This formula will be a
\tconj. If $K_i = \{b_i\}$ with $b_i \in B_R$, we get a formula
$\mbox{NonNull}_{K_i}$ as for the case $L_i = \{b_i\}$ such that $x$ is
replaced by $0$.

If $K_i = K \cdot K'$, then there exists a non null duration in $K_i$ iff
there exists some duration in $K$ and some other in $K'$ and one of them is
non null. Thus $\mbox{NonNull}_{K_i}(\Theta)$ equals
$(\mbox{NonNull}_{K}(\Theta) \wedge \mbox{NonEmpty}_{K'}(\Theta)) \vee
(\mbox{NonEmpty}_{K}(\Theta) \wedge \mbox{NonNull}_{K'}(\Theta))$. If $K_i =
K^+$, then $\mbox{NonNull}_{K_i}(\Theta) = \mbox{NonNull}_{K}(\Theta)$.
Finally, for $L_i = b_i \cdot K_i$, we get the formula
$(\mbox{NonNull}_{\{b_i\}}(x,\Theta) \wedge \mbox{NonEmpty}_{K_i}(\Theta))
\vee (\mbox{NonEmpty}_{\{b_i\}}(x,\Theta) \wedge
\mbox{NonNull}_{K_i}(\Theta))$.
\qed

\begin{lemma}   \label{lem:nonzeno}
  One can construct a ${\sf B}_{x,\Theta}$ formula $\mbox{\rm
    NonZeno}_{L_i}(x,\theta)$ such that for any valuation $v$ and any clock
  value $x_0$, $\mbox{\rm NonZeno}_{L_i}(x_0,v(\theta))$ is {\sc true} iff
  ${\sf R}_i$ contains runs with arbitrarily large durations.
\end{lemma}

\proof
The proof is again similar.

Suppose $L_i = \{\epsilon\}$, then clearly $\mbox{NonZeno}_{L_i}(x,\Theta) =
\false$. Let $L_i = \{b_i\}$ with $b_i \in B$ and associated formula
$\lambda^i_t \wedge \lambda^i_{\leq} \wedge \lambda^i_x \wedge
\lambda^i_{\theta}$. If $\lambda^i_t$ equals $t = \alpha$ or $t = \alpha - x$,
then $\mbox{NonZeno}_{L_i}(x,\Theta) = \false$. If $\lambda^i_t$ equals $t
\cgeq \alpha$ or $t \cgeq \alpha - x$, then $t$ is arbitrarily large iff
$\lambda^i_{\leq} = \true$. In this case, $\mbox{NonZeno}_{L_i}(x,\Theta) =
\lambda^i_x \wedge \lambda^i_{\theta}$, otherwise
$\mbox{NonZeno}_{L_i}(x,\Theta) = \false$.

\begin{sloppy}
Suppose now that $L_i = b_i \cdot K_i$. We begin to construct a \tconj\ 
$\mbox{NonZeno}_{K_i}(\Theta)$ by induction on $K_i$. If $K_i = \{b_i\}$ with
$b_i \in B_R$, then the formula is as in the case $L_i = \{b_i\}$ with $x$
replaced by $0$. If $K_i = K \cdot K'$, then $\mbox{NonZeno}_{K_i}(\Theta)$
equals $(\mbox{NonZeno}_{K}(\Theta) \wedge \mbox{NonEmpty}_{K'}(\Theta)) \vee
(\mbox{NonEmpty}_{K}(\Theta) \wedge \mbox{NonZeno}_{K'}(\Theta))$. If $K_i =
K^+$, then $K_i$ has arbitrarily large durations iff $K$ contains a non null
duration, that is $\mbox{NonZeno}_{K_i}(\Theta) = \mbox{NonNull}_{K}(\Theta)$.
Thus we get for $L_i = b_i \cdot K_i$ the formula
$$(\mbox{NonZeno}_{\{b_i\}}(x,\Theta) \wedge \mbox{NonEmpty}_{K_i}(\Theta))
\vee (\mbox{NonEmpty}_{\{b_i\}}(x,\Theta) \wedge
\mbox{NonZeno}_{K_i}(\Theta)).$$
\end{sloppy}
\qed

\begin{lemma}   \label{lem:min}
  One can construct a Presburger formula $\mbox{\rm Min}_{L_i}(t,x,\theta)$
  such that for any valuation $v$ and any clock value $x_0$, $\mbox{\rm
    Min}_{L_i}(t_0,x_0,v(\theta))$ is {\sc true} iff $t_0$ is the minimum
  duration of runs of ${\sf R}_i$. This formula is equal to $\mu_t \wedge
  \mu_x \wedge \mu_{\theta}$ such that $\mu_t$ is of the form $t = \alpha$ or
  $t = \alpha - x$, $\mu_x$ is an \xconj\ and $\mu_{\theta}$ is a \tconj.
\end{lemma}

\proof
  In this proof, we have to describe the minimum duration by the variable $t$
  and the constraints on it by $\mu_x$ and $\mu_{\theta}$.

Let $L_i = \{\epsilon\}$, then $\mbox{Min}_{L_i}(t,x,\Theta)$ is equal to $(t
= 0) \wedge (x=0)$ if $q$ is a reset-state, and $(t=0) \wedge {\sf
  I}(q)(x,\Theta)$ otherwise. Let $L_i = \{b_i\}$ with $b_i \in B$. Then
looking at the form of $\lambda^i_t$, the minimum duration equals $\alpha$
($\alpha - x$ resp.) (see (\ref{eq:nonempty}) and the sentence just before).
Therefore formula $\mbox{Min}_{L_i}(t,x,\Theta)$ is equal to
\begin{eqnarray}    \label{eq:min}
&& (t = \alpha) \wedge (\bigwedge_{\beta} \alpha \leq \beta - x) \wedge \lambda^i_x \wedge \lambda^i_{\theta} \\
\nonumber
(&&(t = \alpha - x) \wedge (\bigwedge_{\beta} \alpha \leq \beta) \wedge \lambda^i_x \wedge \lambda^i_{\theta} 
  \quad  \mbox{ resp.)}
\end{eqnarray}

Suppose $L_i = b_i \cdot K_i$. Let us begin to construct formula
$\mbox{Min}_{K_i}(t,\Theta)$ the form of which will be $\mu_t \wedge
\mu_{\theta}$. If $K_i = \{b_i\}$ with $b_i \in B_R$, then
$\mbox{Min}_{K_i}(t,\Theta)$ equals (\ref{eq:min}) with $x$ replaced by $0$.
If $K_i = K \cdot K'$, then the minimum duration in $K_i$ equals the sum of
the minimum durations in $K$ and $K'$. Hence, if $\mbox{Min}_{K}(t,\Theta) =
(t = \alpha) \wedge \mu_{\theta}$ and $\mbox{Min}_{K'} = (t = \alpha') \wedge
\mu'_{\theta}$, then $\mbox{Min}_{K_i}(t,\Theta)$ is equal to $(t = \alpha +
\alpha') \wedge \mu_{\theta} \wedge \mu'_{\theta}$. If $K_i = K^+$, then the
minimum duration in $K_i$ is the minimum duration in $K$, i.e.
$\mbox{Min}_{K_i}(t,\Theta) = \mbox{Min}_{K}(t,\Theta)$. Let us come back to
$L_i = b_i \cdot K_i$. Let $\mbox{Min}_{\{b_i\}}(t,x,\Theta)$ be equal to
(\ref{eq:min}) and $\mbox{Min}_{K_i}(t,\Theta)$ be equal $(t = \alpha') \wedge
\mu_{\theta}$. Then $\mbox{Min}_{L_i}(t,\Theta)$ is equal to $(t = \alpha +
\alpha') \wedge (\bigwedge_{\beta} \alpha \leq \beta - x) \wedge \lambda^i_x
\wedge \lambda^i_{\theta} \wedge \mu_{\theta}$ (resp. $(t = \alpha + \alpha' -
x ) \wedge (\bigwedge_{\beta} \alpha \leq \beta ) \wedge \lambda^i_x \wedge
\lambda^i_{\theta} \wedge \mu_{\theta}$) .
\qed

In the next lemma, we are going to construct a formula
$\mbox{Max}_{L_i}(t,x,\Theta)$ that describes the maximum duration $t$ in
$L_i$. Note that durations $t$ in $L_i$ can be arbitrarily large (see Lemma
\ref{lem:nonzeno}). We will thus denote symbolically by $t = \infty$ the (non
existing) maximum duration.

\begin{lemma}   \label{lem:max}
  One can construct a formula $\mbox{\rm Max}_{L_i}(t,x,\theta)$ such that for
  any valuation $v$ and any clock value $x_0$, $\mbox{\rm
    Max}_{L_i}(t_0,x_0,v(\theta))$ is {\sc true} iff $t_0$ is the maximum
  duration of runs of ${\sf R}_i$. This formula is equal to a disjunction of
  formulae $M_t \wedge M_x \wedge M_{\theta}$ such that $M_t$ is of the form
  $t = \alpha$, $t = \alpha - x$ or $t = \infty$, $M_x$ is an \xconj\ and
  $M_{\theta}$ is a \tconj.
\end{lemma}

\proof
  If $L_i = \{\epsilon\}$, then $\mbox{Max}_{L_i}$ is $(t = 0) \wedge (x = 0)$
  if $q$ is a reset-state, and to $(t = 0) \wedge {\sf I}(q)(x,\Theta)$
  otherwise. Let $L_i = \{b_i\}$ with $b_i \in B$. Let us study $\lambda^i_t$
  and $\lambda^i_{\leq}$ equal to $\bigwedge_{\beta} (t \leq \beta - x)$. If
  $\lambda^i_t$ is $t = \alpha$, then $\mbox{Max}_L(t,x,\Theta)$ equals
  $\lambda^i_t \wedge \bigwedge_{\beta} (\alpha \leq \beta - x) \wedge
  \lambda^i_x \wedge \lambda^i_{\theta}$. A similar formula holds when
  $\lambda^i_t$ equals $t = \alpha - x$. If $\lambda^i_t$ is $t \cgeq \alpha$
  with $\lambda^i_{\leq} = \true$, then $\mbox{Max}_L(t,x,\Theta)$ equals $(t
  = \infty) \wedge \lambda^i_x \wedge \lambda^i_{\theta}$. Suppose that
  $\lambda^i_t$ is $t \cgeq \alpha$ with $\lambda^i_{\leq}$ being a non empty
  conjunction $\bigwedge_{\beta} (t \leq \beta - x)$. Then the maximum
  duration is the greatest value $\alpha + ay$, for some $y \in \N$, which is
  less than or equal to the smallest among the $\beta - x$'s, denoted by
  $\beta' - x$.  Assume that $\beta' - x \equiv b \bmod a$ and $\alpha \equiv
  c \bmod a$ for some $b, c \in \{0,\cdots,a-1\}$. If $b \geq c$, then the
  maximum duration is given by formula $M_t$ equal to $t = \beta' - x - (b-c)$
  under the condition $m_{\theta}$ equal to $t \geq \alpha$, i.e. $\beta' - x
  - (b-c) \geq \alpha$ . If $b < c$, then $M_t$ equals $t = \beta' - x -
  (a+b-c)$ under the condition $m_{\theta}$ equal to $\beta' - x - (a+b-c)
  \geq \alpha$. Thus $\mbox{Max}_L(t,x,\Theta)$ is a disjunction over the
  different possible values of $\beta', b$ and $c$ of formulae
$$M_t \wedge m_{\theta} \wedge \lambda_{\theta} \wedge M_{\beta',x,b,c}$$
such
that $M_{\beta',b,c}$ is the conjunction
$$(\bigwedge_{\beta} \beta' \leq \beta) \wedge (\beta' - x \cgeq b) \wedge
(\alpha \cgeq c).$$
A similar argument can be done when $\lambda^i_t$ is $t
\cgeq \alpha - x$.

Let $L_i = b_i \cdot K_i$. Let us first construct $\mbox{Max}_{K_i}$. This
formula will contain no $M_x$. If $K_i = \{b_i\}$ with $b_i \in B_R$, then all
the proof done before for $L_i = \{b_i\}$ can be repeated with $x$ replaced by
$0$. Suppose that $K_i = K \cdot K'$ and that $\mbox{Max}_{K}(t,\Theta)$ and
$\mbox{Max}_{K'}(t,\Theta)$ are a disjunction of formulae $M_t \wedge
M_{\theta}$ and $M'_t \wedge M'_{\theta}$ respectively. If $M_t = (t =
\alpha)$ and $M'_t = (t = \alpha')$, then $\mbox{Max}_{K_i}(t,\Theta)$
contains the conjunction $(t = \alpha + \alpha') \wedge M_{\theta} \wedge
M'_{\theta}$. If $M_t = (t = \infty)$ or $M'_t = (t = \infty)$, then
$\mbox{Max}_{K_i}(t,\Theta)$ contains the conjunction $(t = \infty) \wedge
M_{\theta} \wedge M'_{\theta}$. Suppose that $K_i = K^+$, then the maximum
duration equals $\infty$ if $L$ contains a non null duration (see Lemma
\ref{lem:nonnull}), and $0$ otherwise. Thus $\mbox{Max}_{K_i}(t,\Theta)$ is
the formula $((t = \infty) \wedge \mbox{NonNull}_K(\Theta)) \vee ((t = 0)
\wedge \neg\mbox{NonNull}_K(\Theta))$. Formula $\mbox{Max}_{L_i}(t,x,\Theta)$
for $L_i = b_i \cdot K_i$ can be easily constructed (as done before for $K
\cdot K'$).
\qed

\medskip
\proof (of Proposition \ref{prop:progress}).
  Let us prove that one can construct a ${\sf B}_{x,\Theta}$ formula
  $\Prog_q(x,\Theta)$ such that for any valuation $v$ and any clock value
  $x_0$, $\Prog_q(x_0,v(\Theta))$ is {\sc true} iff there exists an infinite
  run in ${\sf A}^v$ starting with $(q,x_0)$. Such a run exists iff for some
  $q' \in Q$, there exist runs in ${\sf R}_{q,q'}({\sf A}^v,x_0)$ with
  arbitrarily large durations. As ${\sf R}_{q,q'}({\sf A}^v,x_0) = \bigcup_i
  {\sf R}_i$, this is equivalent to say that some ${\sf R}_i$ contains runs
  with arbitrarily large durations. By Lemma \ref{lem:nonzeno}, it follows
  that formula $\Prog_q(x,\Theta)$ is equal to $\bigvee_{q' \in Q} \bigvee_i
  \mbox{NonZeno}_{L_i}(x,\Theta)$.
\qed

\proof (of Proposition \ref{prop:lambda}). Let $\gamma$ be a linear term
  and $\sim \;\in \{<,\leq,>,\geq\}$. We have to show that there exists a
  ${\sf B}_{x,\Theta}$ formula $\llambda_{q,q'}^{\sim\gamma}(x,\Theta)$ such
  that for any valuation $v$ and any clock value $x_0$,
  $\llambda_{q,q'}^{\sim\gamma}(x_0,v(\Theta))$ is {\sc true} iff there exists
  a run in ${\sf R}_{q,q'}({\sf A}^v,x_0)$ with duration $t \sim v(\gamma)$.

  (1) We begin with $\sim \; \in \{<, \leq\}$. To test if there exists a run
  in ${\sf R}_{q,q'}({\sf A}^v,x_0)$ with duration $t \sim v(\gamma)$ is
  equivalent to test that $t_{min} \sim v(\gamma)$ with $t_{min}$ being the
  minimum duration of runs in ${\sf R}_{q,q'}({\sf A}^v,x_0)$. By Lemma
  \ref{lem:min}, the minimum duration for each ${\sf R}_i$ is expressed by
  formula $\mbox{Min}_{L_i}(t,x,\Theta)$. This formula is of the form $\mu_t
  \wedge \mu_x \wedge \mu_{\theta}$ with $\mu_t$ equal to $t = \alpha$ or $t =
  \alpha - x$. Therefore $\llambda_{q,q'}^{\sim\gamma}(x,\Theta)$ is equal to
  $\bigvee_i\llambda_i$, where each $\llambda_i$ is obtained by modifying
  $\mbox{Min}_{L_i}$ as follows: any formula $\mu_t$ equal to $t = \alpha$ ($t
  = \alpha - x$ resp.) is replaced by formula $\alpha \sim \gamma$ ($\alpha -
  x \sim \gamma$ resp.).

(2) We now turn to $\sim \; \in \{>, \geq\}$. The approach is similar but with
the maximum (instead of minimum) duration. By Lemma \ref{lem:max}, the maximum
duration for each ${\sf R}_i$ is expressed by formula
$\mbox{Max}_{L_i}(t,x,\Theta)$. This formula is a disjunction of formulae $M_t
\wedge M_x \wedge M_{\theta}$ with $M_t$ equal to $t = \alpha$, $t = \alpha -
x$ or $t = \infty$. It follows that $\llambda_{q,q'}^{\sim\gamma}(x,\Theta)$
is equal to $\bigvee_i\llambda_i$, where each $\llambda_i$ is obtained by
modifying $\mbox{Max}_{L_i}$ in the following way. If $M_t$ equals $t =
\alpha$, $t = \alpha - x$ or $t = \infty$, then it is replaced by formula
$\alpha \sim \gamma$, $\alpha - x \sim \gamma$ or $\true$ respectively.
\qed

\section{Conclusion}

In this paper, we have completely studied the model-checking problem and the
parameter synthesis problem of the logic \logic, an extension of TCTL with
parameters, over one parametric clock discrete-timed automata. On the negative
side, we showed that the model-checking problem is undecidable. The
undecidability result needs equality in the logic. On the positive side, we
showed that for the fragment F-\logic\ where the equality is not allowed, the
model-checking problem becomes decidable and the parameter synthesis problem
is solvable. Our algorithm is based on automata theoretic principles and an
extension of our method (see \cite{VJFstacs03}) to express durations of runs
of a timed automaton using Presburger arithmetic. With this approach, the
model-checking problem and the parameter synthesis problem are syntactically
translated into Presburger arithmetic which has a decidable theory and an
effective quantifier elimination. The model checking problem is translated
into a Presburger sentence inside which the Presburger decidability process
looks for semantic inconsistencies between the parameters and the parametric
clock. The parameter synthesis problem asks for which values of the parameters
is a F-\logic\ formula true at a given configuration of the timed automaton.
Thanks to Presburger quantifier elimination, this problem is solved by
expressing the values of the parameters in terms of the operations $+$, $\leq$
and $\equiv \bmod a$, $a \in \N^+$.

To the best of our knowledge, this is the first work that studies the
model-checking and parameter synthesis problems with parameters both in the
model (timed automaton) and in the property (\logic\ formula). The problems
solved in this paper are important as it is very natural to refer in the
properties of the system to parameters appearing in the model of the system.
We illustrated in the introduction the kind of properties that can be
expressed and automatically verified in our framework.

Future works could be the following ones. A first work is to give the precise
bordeline between decidability and undecidability.  Is the model-checking
decidable for the logic \logic\ such that equality is forbidden in the
operators $\EU_{\sim \alpha}$ and $\AU_{\sim \alpha}$?  No complexities
issues are given in this paper and only the discrete time is considered.
Presburger theory is decidable with the high {\sc 3ExpTime} complexity. More
efficient algorithms should be designed for particular fragments of F-\logic.
The extension to dense timed models of the method proposed in this paper
should be investigated.

\bibliographystyle{plain}

\end{document}